\documentclass[letter,11pt]{article}
\usepackage[top=1in,bottom=1in,left=1in,right=1in]{geometry}
\usepackage{slashed}
\usepackage{epsfig}
\usepackage{comment}
\usepackage{cancel}
\usepackage{bbm}
\usepackage{array}
\usepackage{bigints}
\usepackage{booktabs}
\usepackage{color}
\usepackage{dsfont}
\usepackage{float}
\usepackage{framed}
\usepackage{graphicx}
\usepackage{indentfirst}
\usepackage{mathrsfs}
\usepackage{multirow}
\usepackage{subdepth}
\usepackage{titlesec}
\usepackage[dotinlabels]{titletoc}
\usepackage{wrapfig}
\usepackage[all]{xy}
\usepackage[vcentermath]{youngtab}
\usepackage{relsize}
\usepackage{hyperref}
\usepackage{xcolor}
\usepackage{soul}
\numberwithin{equation}{section}

\usepackage[utf8]{inputenc}
\usepackage{slashed}
\usepackage{amsmath}
\usepackage{amsfonts}
\usepackage{amssymb}
\usepackage{cite}
\usepackage{mathtools}

\usepackage{cleveref}
\crefname{section}{§}{§§}
\crefname{section}{§}{§§}

   \makeatletter
  \let\over=\@@over \let\overwithdelims=\@@overwithdelims
  \let\atop=\@@atop \let\atopwithdelims=\@@atopwithdelims
  \let\above=\@@above \let\abovewithdelims=\@@abovewithdelims
\renewcommand\section{\@startsection {section}{1}{\z@}%
{-3.5ex \@plus -1ex \@minus -.2ex}%nn
{2.3ex \@plus.2ex}%
{\normalfont\large\bfseries}}

\renewcommand\subsection{\@startsection{subsection}{2}{\z@}%
{-3.25ex\@plus -1ex \@minus -.2ex}%
{1.5ex \@plus .2ex}%
{\normalfont\bfseries}}

\makeatletter
\DeclareFontFamily{OMX}{MnSymbolE}{}
\DeclareSymbolFont{MnLargeSymbols}{OMX}{MnSymbolE}{m}{n}
\SetSymbolFont{MnLargeSymbols}{bold}{OMX}{MnSymbolE}{b}{n}
\DeclareFontShape{OMX}{MnSymbolE}{m}{n}{
    <-6>  MnSymbolE5
   <6-7>  MnSymbolE6
   <7-8>  MnSymbolE7
   <8-9>  MnSymbolE8
   <9-10> MnSymbolE9
  <10-12> MnSymbolE10
  <12->   MnSymbolE12
}{}
\DeclareFontShape{OMX}{MnSymbolE}{b}{n}{
    <-6>  MnSymbolE-Bold5
   <6-7>  MnSymbolE-Bold6
   <7-8>  MnSymbolE-Bold7
   <8-9>  MnSymbolE-Bold8
   <9-10> MnSymbolE-Bold9
  <10-12> MnSymbolE-Bold10
  <12->   MnSymbolE-Bold12
}{}

\let\llangle\@undefined
\let\rrangle\@undefined
\DeclareMathDelimiter{\llangle}{\mathopen}%
                     {MnLargeSymbols}{'164}{MnLargeSymbols}{'164}
\DeclareMathDelimiter{\rrangle}{\mathclose}%
                     {MnLargeSymbols}{'171}{MnLargeSymbols}{'171}
\makeatother

\begin{document}
\begin{titlepage}
\unitlength = 1mm
\ \\
\vskip 3cm
\begin{center}

{\LARGE{\textsc{Shadow Celestial Operator Product Expansions }}}

\vspace{1.25cm}
Elizabeth Himwich$^{*}$ and Monica Pate$^{ \dagger}$

\vspace{.5cm}

$^*${\it  Princeton Center for Theoretical Science, Princeton University, Princeton, NJ 08544}\\ 
$^\dagger${\it  The Center for Cosmology and Particle Physics, New York University, New York, NY 10003}\\ 

\vspace{0.8cm}

\begin{abstract} 
The linearized massless wave equation in four-dimensional asymptotically flat spacetimes is known to admit two families of solutions that transform in highest-weight representations of the Lorentz group ${\rm SL}(2, \mathbb{C})$. The two families are related to each other by a two-dimensional shadow transformation. The scattering states of one family are constructed from standard momentum eigenstates by a Mellin transformation with respect to energy. Their operator product expansion (OPE) is directly related to collinear limits of momentum space amplitudes.  The scattering states of the other family are a priori non-local on the celestial sphere and lack a standard notion of OPE. Such states appear naturally in the context of asymptotic symmetries, but their properties as operators remain largely unexplored. Here we initiate a study, to be continued in a forthcoming companion paper, of a definition of an OPE for shadow operators. We present a useful technical ingredient: the transformation of the OPE coefficients associated to collinear limits under a shadow. Our results can be used to find the coefficients of all three-point functions involving any combination of celestial and shadow primaries. An OPE block is used to account for the contribution from a primary together with its global conformal descendants, all of which contribute when deriving the shadowed OPE coefficients.   Applications involving $U(1)$ currents and stress tensors as well as a chiral current algebra of soft gluons are discussed. 
\end{abstract}

\vspace{1.0cm}
\end{center}

\end{titlepage}

\pagestyle{empty}
\pagestyle{plain}

\def\vx{{\vec x}}
\def\p{\partial}
\def\po{$\cal P_O$}
\def\i{{\rm initial}}
\def\f{{\rm final}}

\pagenumbering{arabic}
 
%%%%%%%%%%%%%%%%---------------------END OF TITLE PAGE AND ABSTRACT---------------------%%%%%%%%%%%%%%%%%%%%%%

\tableofcontents

\section{Introduction}

It is open question whether the holographic principle extends to quantum gravity in asymptotically flat spacetime.  One avenue for progress begins with the simple observation that the Lorentz group in four dimensions ${\rm SO}(3, 1)$ is isomorphic to the global conformal group in two dimensions ${\rm SL}(2, \mathbb{C})$.  While this fact alone may seem too elementary to motivate a holographic correspondence, the symmetry group of four-dimensional asymptotically flat spacetimes remarkably also extends to include the infinite-dimensional Virasoro symmetry of a two-dimensional conformal field theory \cite{Barnich:2009se,Barnich:2010eb,Barnich:2011mi,Kapec:2014opa}.  Although these symmetries are still not sufficient to guarantee a holographic correspondence, the proposal that a two-dimensional conformal field theory (2D CFT) is dual to four-dimensional asymptotically flat gravity has gained significant traction in recent years.  In particular, the highly constrained nature of 2D CFTs --- namely their full specification  by a consistent spectrum of primary field operators and operator product expansion (OPE) coefficients --- makes them attractive candidates for holographic duals.  

To evaluate the merit of this holographic proposal, it is helpful to recast the gravitational scattering problem in a form that manifests the underlying 2D conformal structure. The quantum gravitational scattering problem is formally solved by scattering amplitudes between asymptotic states comprised of asymptotic particles.  Asymptotic particles are typically  defined as irreducible representations of the Poincar\'e group and typically transform in induced representations that diagonalize the four spacetime translations, also known as momentum states.\footnote{One can also consider irreducible representations of the BMS group, known as BMS particles, or extensions thereof.  See e.g.~\cite{Mccarthy:1972ry} and subsequent references for the original treatment of BMS particles, \cite{Bekaert:2024jxs,Bekaert:2025kjb} for a recent treatment of BMS particles, and \cite{Ahmad:2025new} for a novel extension to generalized BMS including a superrotation subgroup. We focus on the ${\rm SL}(2,\mathbb{C})$ subgroup of BMS/Poincar\'e.} By contrast, primary field operators in standard 2D CFTs transform in highest-weight representations of ${\rm SL}(2, \mathbb{C})$ and are the basic building blocks of correlation functions in these theories. Scattering amplitudes between asymptotic particles in highest-weight representations of ${\rm SL}(2, \mathbb{C})$, known as celestial amplitudes, may therefore admit a more natural and straightforward interpretation in terms of observables of a dual conformal field theory \cite{Pasterski:2016qvg, Pasterski:2017kqt, Pasterski:2017ylz}.  In particular, by construction, celestial amplitudes transform under ${\rm SO}(3,1)$ Lorentz transformations like correlation functions of primary field operators under ${\rm SL}(2,\mathbb{C})$ global conformal transformations.  A growing number of studies (see \cite{Pasterski:2021raf} for a fairly complete list as of its print date) support the proposal that celestial amplitudes, with manifest 2D global conformal covariance, can elucidate the structure of gravitational scattering processes. The hypothesized dual theory with correlation functions that compute the celestial amplitudes is known as celestial conformal field theory (CCFT). 

There is a now-standard systematic procedure for constructing celestial amplitudes, which can be broken down into two basic steps.  This procedure will be summarized briefly now and reviewed in detail in Appendix \ref{appA:celestial-amplitude-review}.  The first step is to solve the linearized bulk wave equation subject to highest-weight conditions.  The resulting solutions $\varphi_{h, \bar{h}}(x; z, \bar{z})$ carry the same labels as primary field operators in a 2D conformal field theory, namely left and right conformal weights $(h, \bar{h})$ and a position $(z, \bar{z})$ on the 2D plane.  Enforcing highest-weight conditions ensures that these solutions transform under the bulk Lorentz group like primary field operators under the global conformal group.  The second step is to perform LSZ using these solutions as wavefunctions.\footnote{One may object to using LSZ in a theory of quantum gravity because correlation functions of local operators are no longer gauge invariant, but the procedure for constructing celestial amplitudes can be easily modified to eliminate any application of the LSZ prescription.  Specifically, by expanding the solutions from the first step in the plane wave basis, one can determine the transformation from the plane wave basis to the highest-weight basis.  This is always possible because the plane waves and the highest-weight  solutions each form a complete basis for the wave equation \cite{Pasterski:2017kqt}, as discussed below. Then, scattering amplitudes in the highest-weight basis can be constructed directly from momentum-space amplitudes by applying this basis transformation.}  The resulting amplitudes will transform like correlation functions of global primary field operators, so we represent them by $\langle \mathcal{O}_1 \cdots \mathcal{O}_n\rangle$, a notation usually reserved for correlation functions.  Under global conformal transformations, coordinates on the plane transform via M\"obius transformations:  
\begin{equation}
    \begin{split}
        z \to z' = \frac{az +b}{cz +d}, \quad \quad \quad 
    \bar{z} \to \bar{z}' = \frac{\bar{a} \bar{z}+\bar{b}}{\bar{c} \bar{z}+\bar{d}}, \quad \quad \quad \left(\begin{matrix}
            a& b\\
            c&d
        \end{matrix}\right) \in {\rm SL}(2, \mathbb{C}),
    \end{split}
\end{equation}
and celestial amplitudes transform as
\begin{equation} \label{amp_trans}
    \begin{split}
        \langle \mathcal{O}_1 &(z_1, \bar{z}_1) \cdots \mathcal{O}_n (z_n, \bar{z}_n) \rangle\\  
        &\to 
        \langle \mathcal{O}'_1 (z'_1, \bar{z}'_1) \cdots \mathcal{O}'_n (z'_n, \bar{z}'_n) \rangle 
         = \Big(\prod_{i=1}^n (cz_i+d)^{2h_i}(\bar{c}\bar{z}_i+\bar{d})^{2\bar{h}_i}\Big)\langle \mathcal{O}_1 (z_1, \bar{z}_1) \cdots \mathcal{O}_n (z_n, \bar{z}_n) \rangle.
    \end{split}
\end{equation}
Here all of the labels of the asymptotic states, apart from the 2D position, are represented by the single index $i$ on each operator $\mathcal{O}_i$. 

The massless wave equation admits two families of solutions subject to highest-weight conditions.  The first family is related to the plane waves by a Mellin transform
\begin{equation} \label{states1-intro}
    \begin{split}
        \varphi_{h, \bar{h}}(x; z, \bar{z}) \sim \int_0^\infty \frac{d \omega}{\omega} \omega^\Delta e^{i p\cdot x},
    \end{split}
\end{equation}
where $\Delta = h + \bar{h}$ is the weight and  $s = h - \bar{h}$ is the two-dimensional spin, to be identified with bulk helicity.  The null momentum $p^\mu$ is parametrized by an energy scale $\omega$ and a point $(z,\bar{z})$ on the celestial sphere: 
\begin{equation} 
	p^\mu =  \eta\omega \hat q^\mu (z, \bar{z}), \quad \quad \quad 
    \hat q^\mu (z, \bar{z}) = \big(1+z\bar{z}, z+\bar{z}, -i(z-\bar{z}), 1-z\bar{z}\big).
\end{equation}
The sign $\eta = \pm 1$ distinguishes between outgoing and incoming momenta.  Note that the wavefunctions \eqref{states1-intro} localize to spatial cross sections of null infinity, so the position $(z,\bar{z})$ of the dual celestial operator is identified with the spatial direction of the null momentum.  The states constructed from \eqref{states1-intro} comprise a simple, physically intuitive collection of highest-weight states, which we will refer to as \emph{celestial primary states} (or operators). For $\Delta = 1+ i \lambda$,  $\lambda \in \mathbb{R}$, the celestial primary states were found to span the space of plane wave states  \cite{Pasterski:2017kqt}.
	
The second family of solutions, which also spans the plane waves for $\Delta = 1+ i \lambda$,  $\lambda \in \mathbb{R}$ \cite{Pasterski:2017kqt}, can be used to construct an alternate highest-weight basis. This second family is related to the first by a shadow transformation\footnote{In (2,2) signature Klein space \cite{Atanasov:2021oyu,Melton:2023bjw,Melton:2023hiq}, a further generalization to light transforms \cite{Kravchuk:2018htv} is available and furnishes another candidate basis of interest \cite{Guevara:2021abz,Strominger:2021lvk,Sharma:2021gcz,Himwich:2021dau,Guevara:2021tvr,Jorge-Diaz:2022dmy}. We defer further discussion of split signature to future work.} 
\begin{equation} \label{branch-2}
    \begin{split}
        \widetilde{\varphi}_{h, \bar{h}}(x; z ,\bar{z})& \sim  \int \frac{d^2 w}{2 \pi } \frac{\varphi_{1-h,1-\bar{h}}(x; w ,w)}{(z-w)^{2h}(\bar{z}-\bar{w})^{2\bar{h}}}. 
    \end{split}
\end{equation} 
Note that in contrast to the first family, the location of the celestial operator $(z, \bar{z})$ no longer coincides with a direction in the bulk spacetime.  We refer to states constructed from this second set as \emph{celestial shadow states} (or operators).  

The shadow transformation is a familiar concept from standard conformal field theory \cite{Ferrara:1972xe,Ferrara:1972ay,Ferrara:1972uq,Ferrara:1972kab,Simmons-Duffin:2012juh}. However, in a standard conformal field theory, the shadow is only a formal (global) conformally covariant operation and in particular \emph{does not} produce another local primary field, despite the fact that it is labeled by a point and transforms covariantly under ${\rm SL}(2, \mathbb{C})$.  Nevertheless, the shadow transformation is still a useful operation, for example in constructing conformal partial waves \cite{Ferrara:1972xe,Ferrara:1972ay,Ferrara:1972uq,Ferrara:1972kab,Simmons-Duffin:2012juh,Gadde:2017sjg,Caron-Huot:2017vep,Simmons-Duffin:2017nub}. On the other hand, from the perspective of celestial conformal field theory, the role and interpretation of the shadow transformation is less clear, and the combination of celestial primary and celestial shadow states that comprise the set of primary field operators of the celestial dual is still under investigation. At the level of highest-weight solutions to the wave equation, both branches of solutions appear on equal footing.  In fact, they are related by bulk inversions, which is a symmetry of the linearized massless wave equation \cite{Jorstad:2023ajr,Chen:2024kuq}.\footnote{In contrast, the massive wave equation is not invariant under bulk inversions and consequently there is only one family of solutions for massive particles. In the massive case, the shadow transformation simply maps between highest-weight solutions within the single solution set.} However, as one might expect, they have distinct properties when appearing in celestial amplitudes.

In general, the celestial primary basis is computationally more tractable and as a result has been studied far more intensively.  A  number of striking universal results have been established in this basis, many of which are concretely related to the identification between 2D positions and 4D directions.  For example, collinear limits in momentum space correspond directly to OPEs of celestial primary operators \cite{Fan:2019emx,Pate:2019lpp,Himwich:2021dau}.  Consequently, the known universal behavior of scattering amplitudes at leading order in the collinear limit guarantees universal behavior of the OPE at leading order.  In fact, universal leading OPE coefficients can be derived fully within the holographic dual by simply augmenting the global ${\rm SL} (2, \mathbb{C})$ symmetry algebra with exotic  charges that  generate bulk translations \cite{Himwich:2021dau}. When restricted to gravitons, the operator algebra remarkably organizes into the infinite-dimensional algebra of ${\rm w}_{1+\infty}$ \cite{Guevara:2021abz,Strominger:2021lvk} and OPEs between gravitons and other celestial primary operators imply that massless particles transform in non-trivial representations of this symmetry \cite{Himwich:2021dau}. In addition, celestial primary operators respect standard Ward identities in the presence of current or stress tensor operator insertions \cite{Strominger:2013lka,He:2015zea,Kapec:2016jld,Cheung:2016iub}.  Specifically, they behave like local operators and local sources of charge in correlation functions involving single spin-two stress-tensor or spin-one current operator  insertions. 

A similar understanding of the shadow primary basis has yet to be established. In practice, this basis is both more technically cumbersome and less physically intuitive, properties which have inhibited significant progress. Given the a priori non-local nature of the shadow basis, one might indeed wonder whether it should play any role in the construction of a holographic dual.  However, a local stress tensor\footnote{Up to modifications needed in the presence of other celestial shadow operators and massive operators, as discussed below and in  \cite{Banerjee:2022wht,Himwich:2023njb}.} that generates conformal transformations on massless celestial primaries is known to correspond to a mode of the graviton in the shadow primary basis  \cite{Kapec:2016jld}.  As such, some treatment of this basis is necessary for the dual to admit a local stress tensor like an ordinary conformal field theory. The celestial shadow basis is also employed in the construction of candidate norms for the dual theory \cite{Fan:2021isc,Crawley:2021ivb,Cotler:2023qwh}.\footnote{\cite{Cotler:2023qwh} considers an integer basis of solutions to the massless Klein-Gordon equation that is complete in the sense of decomposing the Wightman function and involves both primaries and shadows. On the other hand, the discrete basis of \cite{Freidel:2022skz}, which is complete in the sense of reconstructing solutions in Schwartz space, does not involve shadows. Neither does the basis $\Delta\in \mathbb{R}$ considered by the modified contour prescription in \cite{Mitra:2024ugt}.} In particular, the shadow transformation takes the delta-function singularities that arise in two- and three-point functions of operators in the celestial primary basis and transforms them to the standard power-law form \cite{Pasterski:2017kqt,Chang:2022seh}. Along similar lines, the singular behavior of celestial four-point amplitudes due to bulk translation invariance can be resolved by putting one operator in the shadow basis.  This result was established at tree-level in Yang-Mills in \cite{Fan:2021isc}, in massless $\phi^3$ scalar theory in \cite{Fan:2023lky}, for massive and massless scalars interacting via a three-point interaction in \cite{Chang:2022jut},  in Einstein gravity in \cite{Surubaru:2025qhs,Bhattacharyya:2025nfp}, and in Einstein-Yang-Mills in \cite{Liu:2025dhh}.  Shadow celestial operators have appeared in several other contexts including the study of magnetic large gauge symmetry on the celestial sphere \cite{Nande:2017dba}, the flat space limit of AdS/CFT \cite{Hijano:2020szl,Duary:2022pyv,deGioia:2023cbd}, new prescriptions for celestial observables \cite{Sleight:2023ojm,Jorstad:2023ajr,Jain:2023fxc,Banerjee:2024yir}, the construction of current and stress tensor operators in higher dimensions \cite{Kapec:2017gsg,Kapec:2021eug}, and parallel transport in the moduli space of vacua \cite{Kapec:2022hih,Kapec:2022axw,Narayanan:2024qgb}.  Despite this list, in practice it is rather cumbersome to work with the shadow celestial basis and so general results and explicit calculations are more limited. 

Given the evidence favoring the inclusion of at least some celestial shadow states, it is useful to establish   their behavior as operators and some of their universal properties, in analogy with celestial primary states. The operator product expansion is a powerful tool in conformal field theories.  As discussed above, in celestial holography, the OPE coefficients of the celestial primary operators admit a simple, universal bulk interpretation in terms of collinear limits and have proven remarkably robust.  The holomorphic/antiholomorphic OPE coefficients naturally arise in the self-dual theories \cite{Monteiro:2011pc,Adamo:2021lrv,Ball:2021tmb,Kmec:2024nmu}, but take the same form at tree level in minimally coupled gravitational and gauge theories in flat space \cite{Guevara:2021abz,Strominger:2021lvk}, in certain non-trivial backgrounds \cite{Costello:2022wso,Adamo:2023zeh}, and even under variations of the holographic dictionary (such as leaf amplitudes \cite{Melton:2023bjw,Melton:2024jyq}).

If shadowed states are to be included in the dual theory, it is essential to understand whether there is a prescription that allows them to be treated as operators that are more familiar  in a standard 2D CFT, in particular as operators with a meaningful notion of OPE.  An obvious challenge is that the integral \eqref{branch-2} in the construction of celestial shadow states obscures a direct relation to collinear limits in momentum space, and more generally, because it is a non-local transformation, appears to inhibit a discussion of OPEs.  Here, we initiate a study aimed to address this issue.  

Our first step will be to establish a covariant form of the shadow transform of the OPE coefficients associated to collinear limits. In the case of ordinary celestial primary states, OPE coefficients of celestial primary  operators could be derived by applying a basis transformation (i.e.~a Mellin transform) to only the leading term in a collinear expansion in momentum space. Subleading terms associated to that conformal family (i.e.~descendants) are then fully fixed by the two-dimensional global conformal symmetry. As a non-local integral transformation over the sphere, the shadow does not preserve the coincident collinear limit. Hence there is no reason for the leading term in the collinear limit to dominate over all other terms when performing the shadow transformation, so one must also consider subleading terms when calculating the shadowed OPE coefficients.  However, since the shadow transformation just reorganizes a given conformal family, a primary will only mix with its global conformal descendants under the shadow transform.  Therefore, it is natural to shadow-transform the OPE contribution of a single primary together with its infinite tower of global conformal descendants, rather than the entire operator product expansion involving multiple conformal families. To facilitate this calculation, we use the ``OPE block'' \cite{Czech:2016xec,CarneirodaCunha:2016zmi,deBoer:2016pqk}, which succinctly captures the requisite terms by an integral formula that then can be readily shadow-transformed.\footnote{OPE blocks have been mentioned in the context of celestial CFT in \cite{Pate:2020notes,Guevara:2021tvr,Kulp:2024scx}.} Our result can be used to find the coefficients of all three-point functions involving any combination of celestial and shadow primaries. 

The next step, which will be presented in an upcoming companion paper \cite{Himwich:2025new}, is to analyze the contribution of shadowed OPE blocks in higher-point correlation functions.   Shadowing the OPE block explicitly takes the original operator outside the radius of convergence of its OPE, namely the region away from other operator insertions in a correlation function. Thus it is expected that the (possibly non-local) analytic structure in correlation functions of shadowed operators is not captured by the shadowed OPE blocks alone. In addition, in a standard CFT, the OPE block of local operators reproduces the contribution of a global primary and its global descendants in a correlation function only after a particular projection is performed \cite{Simmons-Duffin:2012juh}. Our upcoming work will investigate the analytic structure of correlation functions with shadowed operators to determine how shadowed celestial states can be formulated as (possibly non-local) operators in the dual theory.\footnote{For instance, non-local operators attached to defect lines are perfectly standard in 2D CFT, see for instance \cite{Chang:2018iay} for an extensive recent treatment.} Such a prescription is crucial in order to make sense of the celestial stress tensor OPE and  appears likely to involve a modification of the OPE block and/or shadow integral kernel, which has been explored from a different perspective in \cite{Banerjee:2022wht}. 

The paper is organized as follows. In Section \ref{sec:2}, we review the shadow transformation, introduce the OPE block and use it to derive general formulas for various shadows of OPE coefficients.  In Section \ref{sec:3}, several applications of the results from the previous section are studied. In Subsection \ref{sec:3.1}, starting with the OPE between a $U(1)$ current and a charged operator, we derive the shadow of the OPE coefficient corresponding to the shadow of the charged operator. We find that the net effect of the shadow is to flip the sign of the OPE coefficient. In Subsection \ref{sec:st}, we perform a similar analysis for the OPE between a celestial shadow operator and a stress tensor.  Again, we find that the result differs from the standard stress tensor OPE coefficient by a sign. These signs are a reflection of the non-locality of shadowed operators, and indicate that the analytic contribution of shadowed operators to correlation functions involves more than an  operator product expansion with the shadowed OPE coefficients. In Subsection \ref{sec:3.2}, we turn to an application in non-abelian gauge theory. We close with a discussion of other potentially interesting applications that are not studied here and comments on our future investigation of shadowed operators in correlation functions.  A review of highest-weight solutions to the massless wave equation is provided in Appendix \ref{appA:celestial-amplitude-review}. Useful formulas for various integrals used throughout  are collected in  Appendix \ref{appB}.

\section{Shadows of OPE Coefficients from OPE Blocks} \label{sec:2}

In this section, we first review the shadow transformation in Subsection \ref{sec:ShadowDef} and introduce the OPE block in Subsection \ref{sec:OPEblock}. Then, we use the OPE block to derive general formulas for various shadows of OPE coefficients in Subsection \ref{sec:OPEcoefs}. A comparison of our results with those in the literature is provided in Subsection \ref{sec:24}.

\subsection{Shadow Transformations} \label{sec:ShadowDef}

The shadow operation
\begin{equation}\label{shadowdef}
    \begin{split}
	\widetilde{\mathcal{O}} (z, \bar{z})
		= \mathcal{R}_{h, \bar{h}} \int \frac{d^2 w}{2 \pi}\frac{1}{(z-w)^{2-2h}(\bar{z}-\bar{w})^{2-2\bar{h}}} ~\mathcal{O} (w, \bar{w}),
    \end{split}
\end{equation}
is an integral transformation of a global conformal primary field operator $\mathcal{O}(z, \bar{z})$ of conformal weight $(h, \bar{h})$ that transforms like a global conformal primary of weight $(1-h, 1- \bar{h})$.  Here $\mathcal{R}_{h, \bar{h}}$ is a normalization factor, which does not affect the transformation of these states under ${\rm SL}(2, \mathbb{C})$.

For subsequent calculations, it will be helpful to notice that we can express the shadow transformation in the following way:
\begin{equation} \label{shadow-2}
    \begin{split}
        \widetilde{\mathcal{O}}_i (z, \bar{z})=\mathcal{R}_i \int \frac{d^2 w}{2 \pi} ~\llangle \widetilde{\mathcal{O}}_i(z, \bar{z}) \widetilde{\mathcal{O}}_i(w, \bar{w}) \rrangle ~\mathcal{O}_i (w, \bar{w}),
    \end{split}
\end{equation}
where the weights are subsumed into the label $i$.  Here, we use $\llangle \cdots \rrangle$ to denote the standard (i.e.~power-law) kinematically-allowed structure for a correlation function between operators of given conformal weights. For example, 
\begin{equation} \label{kinematic-correlator}
    \begin{split}
        \llangle \mathcal{O}_i(z, \bar{z}) \mathcal{O}_i(w, \bar{w}) \rrangle
         & = \frac{1}{(z-w)^{2h_i}(\bar{z}-\bar{w})^{2\bar{h}_i}},\\
       \llangle  \widetilde{\mathcal{O}}_i(z, \bar{z}) \widetilde{\mathcal{O}}_i(w, \bar{w}) \rrangle
       &=\frac{1}{(z-w)^{2-2h_i}(\bar{z}-\bar{w})^{2-2\bar{h}_i}},\\
       \llangle \mathcal{O}_1(z_1, \bar{z}_1) \mathcal{O}_2(z_2, \bar{z}_2)\mathcal{O}_3(z_3, \bar{z}_3) \rrangle   &=
       \frac{1}{z_{12}^{h_1+h_2-h_3}z_{23}^{h_2+h_3-h_1}z_{31}^{h_3+h_1-h_2}} \\& \quad \quad \quad \times\frac{1}{\bar{z}_{12}^{\bar{h}_1+\bar{h}_2-\bar{h}_3}\bar{z}_{23}^{\bar{h}_2+\bar{h}_3-\bar{h}_1}\bar{z}_{31}^{\bar{h}_3+\bar{h}_1-\bar{h}_2}}.
    \end{split}
\end{equation}
In particular, these expressions should be treated as c-number, not operator-valued, expressions. 

The notation in \eqref{shadow-2} makes the transformation under ${\rm SL}(2, \mathbb{C})$ straightforward to check.  Namely, using the transformation of the primary field operator 
\begin{equation} \label{op-transform-sl2c}
    \begin{split}
        \mathcal{O}_i (z, \bar{z}) \to 
        \mathcal{O}'_i (z', \bar{z}')
        = (cz+d)^{2h_i} (\bar{c}\bar{z} + \bar{d})^{2\bar{h}_i} \mathcal{O}_i (z, \bar{z}),
    \end{split}
\end{equation}
the transformation of the kinematic two-point structure
\begin{equation}
    \begin{split}
        \llangle \mathcal{O}_i(z, \bar{z}) \mathcal{O}_i(w, \bar{w}) \rrangle
        \to (cz+d)^{2h_i} (\bar{c}\bar{z} + \bar{d})^{2\bar{h}_i}
        (cw+d)^{2h_i} (\bar{c}\bar{w} + \bar{d})^{2\bar{h}_i}\llangle \mathcal{O}_i(z, \bar{z}) \mathcal{O}_i(w, \bar{w}) \rrangle,
    \end{split}
\end{equation}
and the transformation of the measure
\begin{equation}
    \begin{split}
        d^2w \to d^2 w' = \frac{d^2w}{(cw+d)^2 (\bar{c}\bar{w} + \bar{d})^2},
    \end{split}
\end{equation}
one immediately finds the transformation of the shadow
\begin{equation}
    \begin{split}
        \widetilde{\mathcal{O}}_i (z, \bar{z}) \to 
        \widetilde{\mathcal{O}}'_i (z', \bar{z}')
        = (cz+d)^{2-2h_i} (\bar{c}\bar{z} + \bar{d})^{2-2\bar{h}_i} \widetilde{\mathcal{O}}_i (z, \bar{z}),
    \end{split}
\end{equation}
which precisely matches the transformation of a primary field operator of weights $(1-h_i, 1-\bar{h}_i)$. 

\subsection{Celestial OPEs and OPE Blocks} \label{sec:OPEblock}

The operator product expansion of celestial primary operators for massless particles is determined at leading order in a holomorphic limit by the three-point amplitudes of the bulk theory \cite{Pate:2019lpp,Fan:2019emx,Himwich:2021dau}. At subleading orders, generically both descendants as well as other primary operators contribute. ${\rm SL}(2,\mathbb{C})$ Lorentz covariance implies an operator product expansion of the general form 
\begin{equation} \label{genOPE}
    \begin{split}
	\mathcal{O}_{1}(z_1, \bar{z}_1)&\mathcal{O}_{2}(z_2, \bar{z}_2)
			= \sum_{ \substack{{\rm primaries} \\P}} \frac{C_{12}{}^{P}}{z_{12}^{h_1+h_2 -h_P} \bar{z}_{12}^{\bar{h}_1+\bar{h}_2 -\bar{h}_P}}
			\sum_{n , \bar{n} = 0}^\infty  \beta^{(n,\bar{n})}_{12P} z_{12}^n \bar{z}_{12}^{\bar{n}} \partial_{z_2}^n \partial_{\bar{z}_2}^{\bar{n}} \mathcal{O}_{P}(z_2, \bar{z}_2) 
		.
	\end{split}
\end{equation}
Here the expansion is organized by families of global conformal primary operators labeled by $P$. We take $\beta^{(0,0)}_{12P} = 1$ so that $C_{12}{}^{P}$  is the OPE coefficient of the primary operator $P$. The remaining coefficients $\beta^{(n,\bar{n})}_{12P}$ are fixed by requiring that both sides of \eqref{genOPE} transform identically under ${\rm SL}(2, \mathbb{C})$.  

In the next subsection, we will determine the shadow transformations of OPE coefficients for all combinations of shadows of the operators $\mathcal{O}_{1} $, $\mathcal{O}_{2} $, and $\mathcal{O}_{P}$. Here we introduce the OPE block, which will facilitate our analysis.   

Note that the shadow transformation is manifestly global conformally covariant, so it will not mix contributions from different global conformal primaries. Therefore, we can treat the contribution from each primary individually, which we denote by
\begin{equation}\label{only_P}
    \begin{split}
 	\mathcal{O}_{1}(z_1, \bar{z}_1)\mathcal{O}_{2}(z_2, \bar{z}_2) \Big|_{P}
			\equiv \frac{C_{12}{}^P}{z_{12}^{h_1+h_2 -h_P} \bar{z}_{12}^{\bar{h}_1+\bar{h}_2 -\bar{h}_P}}
			\sum_{n , \bar{n} = 0}^\infty  \beta^{(n,\bar{n})}_{12P} z_{12}^n \bar{z}_{12}^{\bar{n}} \partial_{z_2}^n \partial_{\bar{z}_2}^{\bar{n}} \mathcal{O}_{P}(z_2, \bar{z}_2) .
    \end{split}
\end{equation}

On the other hand, the shadow transformation will mix a primary with its descendants, so we will need to include the infinite tower of global conformal descendants associated to the primary $P$.  It is still somewhat cumbersome to work with \eqref{only_P}, especially if we are interested in shadowing any one of the operators $\mathcal{O}_{1} $, $\mathcal{O}_{2} $, or $\mathcal{O}_{P}$.  In order to work simultaneously with a primary and its global conformal descendants, we consider the OPE block\footnote{Note that our definition of the OPE block differs from the one introduced in \cite{Czech:2016xec} by a factor of $z_{12}^{h_1+h_2} \bar{z}_{12}^{\bar{h}_1+\bar{h}_2}$.}
\begin{equation}\label{def_OPEblock}
    \begin{split}
		\mathcal{O}_{1}(z_1, \bar{z}_1)\mathcal{O}_{2}(z_2, \bar{z}_2) \Big|_{P} = C'_{12}{}^P \int \frac{d^2 w}{2 \pi} 
           ~ \llangle \mathcal{O}_1(z_1, \bar{z}_1)\mathcal{O}_2(z_2, \bar{z}_2)\widetilde{\mathcal{O}}_P(w, \bar{w}) \rrangle ~\mathcal{O}_P(w, \bar{w}),
	\end{split}
\end{equation}
where we have employed the kinematic three-point structure in \eqref{kinematic-correlator} with the third operator of weight $(1-h_P, 1-\bar{h}_P)$. Here $C'_{12}{}^P$ is related to $C_{12}{}^P$ by a fixed constant of proportionality, which we will determine shortly. We will now argue that \eqref{def_OPEblock} is an equivalent way of representing the contribution to the $\mathcal{O}_{1}\mathcal{O}_{2}$ OPE of a primary $P$ and its global conformal descendants in three-point functions \cite{Ferrara:1972xe}. In higher-point functions, a projection is required to recover the OPE contribution after inserting the OPE block \cite{Simmons-Duffin:2012juh}. We will discuss this further in our upcoming work.  

To begin, note that the OPE block is manifestly covariant under global conformal transformations. Specifically, under ${\rm SL}(2, \mathbb{C})$ transformations \eqref{op-transform-sl2c}, the right-hand side of \eqref{def_OPEblock} transforms in precisely the same way as the product of operators $\mathcal{O}_{1}$ and $\mathcal{O}_{2}$, and thus includes the contribution from the infinite sum of global descendants of $\mathcal{O}_{P}$ in \eqref{only_P}. 

The sum \eqref{only_P} can be explicitly recovered from \eqref{def_OPEblock} by Taylor-expanding $\mathcal{O}_P(w, \bar{w})$ about $(z_2, \bar{z}_2)$ and performing the integrals in $(w, \bar{w})$.  For example, to extract the OPE coefficient of the leading primary, we only need the leading term in the Taylor expansion:
\begin{equation}
    \begin{split}
        \mathcal{O}_{1}(z_1, \bar{z}_1)\mathcal{O}_{2}(z_2, \bar{z}_2) \Big|_{P} \sim C'_{12}{}^P \int \frac{d^2 w}{2 \pi} ~ \llangle \mathcal{O}_1(z_1, \bar{z}_1)\mathcal{O}_2(z_2, \bar{z}_2)\widetilde{\mathcal{O}}_P(w, \bar{w}) \rrangle ~\mathcal{O}_P(z_2, \bar{z}_2).
    \end{split}
\end{equation}
Comparing with \eqref{only_P}, we find $C'_{12}{}^P$ is related to $C_{12}{}^P$ by
\begin{equation} \label{C12P-integral}
    \begin{split}
        C_{12}{}^P = C'_{12}{}^P \int \frac{d^2 w}{2 \pi} 
           ~ \llangle \mathcal{O}_1(1)\mathcal{O}_2(0)\widetilde{\mathcal{O}}_P(w, \bar{w}) \rrangle.
    \end{split}
\end{equation}

To clarify the scope of our analysis, it is also helpful to specify the relation of $C_{12}{}^P$ to the coefficient of the three-point function.  For this purpose, first note that \eqref{C12P-integral} can equivalently be written as 
\begin{equation} \label{eq:C12Pblock} 
    \begin{split}
        C_{12}{}^P 
           &=  C'_{12}{}^P \int \frac{d^2 w}{2 \pi} 
           ~ \lim_{z, \bar{z} \to \infty}\frac{\langle \mathcal{O}_P(z, \bar{z})\mathcal{O}_P(w, \bar{w}) \rangle}{\langle \mathcal{O}_P(z, \bar{z})\mathcal{O}_P(0) \rangle} \llangle \widetilde{\mathcal{O}}_P(w, \bar{w})\mathcal{O}_1(1)\mathcal{O}_2(0) \rrangle,
    \end{split}
\end{equation}
where this expression involves the full two-point function (as opposed to the kinematic structure) and holds regardless of the normalization of $\mathcal{O}_P$. Also note that, assuming  an orthogonal basis of primaries, the OPE coefficient is directly related to the $\mathcal{O}_P$-normalization-independent expression for the canonically normalized three-point function coefficient $C_{12P}$:
\begin{equation} \label{eq:C12Pratio}
    \begin{split}
         C_{12}{}^P = C_{12P} \equiv \lim_{z, \bar{z} \to \infty} \frac{\langle \mathcal{O}_P(z, \bar{z})\mathcal{O}_1(1) \mathcal{O}_2(0)  \rangle}{\langle \mathcal{O}_P(z, \bar{z})\mathcal{O}_P (0)\rangle}. 
    \end{split}
\end{equation}
From the equality of \eqref{eq:C12Pblock} and \eqref{eq:C12Pratio}, we observe that the contribution of the OPE block \eqref{def_OPEblock} exactly reproduces the analytic structure of the OPE contribution to the three-point function (and the corresponding three-point coefficients) without performing any sort of projection.  As a direct consequence, the shadow transformations of the OPE coefficients that we derive in this paper will also apply to the coefficients of the canonically normalized three-point functions with shadowed operators.

Assuming that $h_i- \bar{h}_i \in \mathbb{Z}$,\footnote{We do not keep track of phases that likely arise for fermions, which have $h_i- \bar{h}_i \in\mathbb{Z}+ \frac{1}{2}$.} the integral \eqref{C12P-integral} can be performed using \eqref{2point_gen}
and we find 
\begin{equation} \label{C-norm}
    \begin{split}
        C_{12}{}^{P}
           & = C'_{12}{}^P  ~ \frac{\Gamma(h_P-h_{12})\Gamma(h_P+h_{12})\Gamma(1-2h_P)}{\Gamma(1-\bar{h}_P+\bar{h}_{12})\Gamma(1-\bar{h}_P-\bar{h}_{12})\Gamma(2\bar{h}_P)},
    \end{split}
\end{equation} 
where  $h_{12} = h_1-h_2$. Expanding \eqref{def_OPEblock} to subleading orders, we likewise recover explicit expressions for the coefficients $\beta^{(n, \bar{n})}_{12P}$:
\begin{equation}\label{betamn1}
	\begin{split}
		\beta^{(n,\bar{n})}_{12P} 
	&=  \frac{(-1)^{n-\bar{n}}}{n!\bar{n}!} \frac{ \Gamma(h_P+h_{12} +n)\Gamma(1-2h_P-n)\Gamma(1-\bar{h}_P-\bar{h}_{12} )\Gamma(2\bar{h}_P)} {\Gamma(h_P+h_{12})\Gamma(1-2h_P   )\Gamma(1-\bar{h}_P-\bar{h}_{12} -\bar{n})\Gamma(2\bar{h}_P+\bar{n})}.
    \end{split}
\end{equation}
With the relation between $C'_{12}{}^P$ and the primary OPE coefficients at hand, we can now readily shadow transform the OPE block to obtain the shadow of OPE coefficients.  
		 
\subsection{Shadowing the OPE Block} \label{sec:OPEcoefs}
	
In this subsection, we determine the relation between OPE coefficients and their shadows.  To do so, we study the transformation of the OPE block in \eqref{def_OPEblock} under shadow transformations of each of the different operators.

First, we determine the result of shadow-transforming $\mathcal{O}_{1}$ in \eqref{def_OPEblock}.  Using the definition of the shadow \eqref{shadowdef}, we find   
\begin{equation}
    \begin{split}
	\widetilde{\mathcal{O}}_{1}(z_1,& \bar{z}_1) \mathcal{O}_{2}(z_2, \bar{z}_2) \Big|_{P}\\
	&= \mathcal{R}_1 C'_{12}{}^P \int \frac{d^2 z}{2 \pi}\frac{d^2 w}{2 \pi} ~\llangle \widetilde{\mathcal{O}}_1(z_1, \bar{z}_1) \widetilde{\mathcal{O}}_1(z, \bar{z}) \rrangle  
    \llangle \mathcal{O}_1(z, \bar{z})\mathcal{O}_2(z_2, \bar{z}_2)\widetilde{\mathcal{O}}_P(w, \bar{w}) \rrangle  \mathcal{O}_P(w, \bar{w}).
    \end{split}
\end{equation}
The integral over $z$ can be evaluated using the identity \eqref{identity3} and the result can be put the form
\begin{equation}
	\begin{split}
		\widetilde{\mathcal{O}}_{1}(z_1,& \bar{z}_1)  \mathcal{O}_{2}(z_2, \bar{z}_2) \Big|_{P}\\
		 &= \mathcal{R}_1 C'_{12}{}^P 
            (-1)^{h_{12}+h_P- \bar{h}_{12}-\bar{h}_P}
            \frac{\Gamma (2 h_1-1)}{\Gamma (2-2 \bar{h}_1)}
            \frac{\Gamma
   (2-h_1-h_2-h_P) \Gamma
   (h_P-h_{12})}{ \Gamma
   (\bar{h}_1+\bar{h}_2+\bar{h}_P-1)\Gamma
   (\bar{h}_{12}-\bar{h}_P+1)}\\& \quad \quad \times\int \frac{d^2 w}{2 \pi} ~\llangle \widetilde{\mathcal{O}}_1(z_1, \bar{z}_1)\mathcal{O}_2(z_2, \bar{z}_2)\widetilde{\mathcal{O}}_P(w, \bar{w}) \rrangle  \mathcal{O}_P(w, \bar{w}).
	\end{split}
\end{equation}
Since this precisely matches the form of an OPE block involving a shadowed operator, we can immediate read off $C'_{\tilde{1}2}{}^P$:
\begin{equation} \label{pre-ope-relation}
    \begin{split}
        C'_{\tilde{1}2}{}^{P} = \mathcal{R}_1 C'_{12}{}^P(-1)^{h_{12}+h_P- \bar{h}_{12}-\bar{h}_P}
            \frac{\Gamma (2 h_1-1)}{\Gamma (2-2 \bar{h}_1)}
            \frac{\Gamma
   (2-h_1-h_2-h_P) \Gamma
   (h_P-h_{12})}{ \Gamma
   (\bar{h}_1+\bar{h}_2+\bar{h}_P-1)\Gamma
   (\bar{h}_{12}-\bar{h}_P+1)}.
    \end{split}
\end{equation}
To obtain an explicit expression for the shadowed OPE coefficient in terms of the original OPE coefficient, we use \eqref{C-norm} and find 
\begin{equation}\label{OPEs1}
	\begin{split}
		C_{\tilde12}{}^{P}
			& = C_{12}{}^P\mathcal{R}_1 ~\frac{\Gamma(2h_1-1)}{\Gamma(2-2\bar{h}_1)} \frac{\Gamma(1-\bar{h}_P-\bar{h}_{12})}{\Gamma(h_P+h_{12})} \frac{\Gamma(h_P+1-h_1-h_2)}{\Gamma(\bar{h}_1+\bar{h}_2-\bar{h}_P)}.
	\end{split}
\end{equation}
		
Thus far we have kept  normalization $\mathcal{R}$ generic since it does not affect  global conformal transformation properties.   One  common choice for $\mathcal{R}$ is  motivated by the observation
\begin{equation}\label{shadowinverse}
	\begin{split}
		\int \frac{d^2 z_2}{2\pi}~\llangle \mathcal{O}_i(z_1, \bar{z}_1) \mathcal{O}_i(z_2, \bar{z}_2) \rrangle  ~\widetilde{\mathcal{O}}_{i}(z_2, \bar{z}_2)
		= \mathcal{R}_i \frac{\Gamma (2h_i-1)\Gamma(1-2h_i)  }{  \Gamma (2-2\bar{h}_i)\Gamma (2 \bar{h}_i ) } \mathcal{O}_{i}(z_1, \bar{z}_1),
	\end{split}
\end{equation}
which can be derived by using the expression \eqref{shadowdef} for the shadow and the identity  \eqref{integral2}.  Note then that by choosing
\begin{equation}\label{def_R}
    \mathcal{R}_i = \frac{\Gamma(2-2 \bar{h}_i)}{\Gamma (2h_i-1)},
\end{equation}
we have the simple relation
\begin{equation}\label{ss1}
\widetilde{\widetilde{\mathcal{O}}}_{i}(z, \bar{z}) = \mathcal{O}_{i}(z, \bar{z}).
\end{equation}
For  the choice \eqref{def_R}  of $\mathcal{R}$, \eqref{OPEs1} simplifies to 
\begin{equation}\label{OPEs1fixed}
    \begin{split}
	C_{\tilde12}{}^{P}
			& = C_{12}{}^P \frac{\Gamma(1-\bar{h}_P-\bar{h}_{12})}{\Gamma(h_P+h_{12})} \frac{\Gamma(h_P+1-h_1-h_2)}{\Gamma(\bar{h}_1+\bar{h}_2-\bar{h}_P)},
	\end{split}  
\end{equation}
which obeys
\begin{equation}
    C_{\tilde {\tilde 1}2}{}^P  = C_{12}{}^P.
\end{equation}		

The result of shadowing $\mathcal{O}_{2}$ can be immediately deduced from \eqref{OPEs1} by simply exchanging the labels $1 \leftrightarrow 2$:
\begin{equation}\label{OPEs2}
    \begin{split}
	C_{1\tilde 2}{}^{P}
		& = C_{12}{}^P \mathcal{R}_2\frac{\Gamma(2h_2-1)}{\Gamma(2-2\bar{h}_2)} \frac{\Gamma(1-\bar{h}_P+\bar{h}_{12})}{\Gamma(h_P-h_{12})} \frac{\Gamma(h_P+1-h_1-h_2)}{\Gamma(\bar{h}_1+\bar{h}_2-\bar{h}_P)}.
    \end{split}
\end{equation}
Again, choosing $\mathcal{R}$ given by \eqref{def_R}, the shadowed OPE coefficient simplifies to
\begin{equation} \label{OPEs2fixed} 
    \begin{split}
	C_{1\tilde 2}{}^{P}
		& = C_{12}{}^P \frac{\Gamma(1-\bar{h}_P+\bar{h}_{12})}{\Gamma(h_P-h_{12})} \frac{\Gamma(h_P+1-h_1-h_2)}{\Gamma(\bar{h}_1+\bar{h}_2-\bar{h}_P)},
    \end{split}
\end{equation}
which likewise satisfies $C_{1 \tilde{\tilde 2}}{}^{P} = C_{12}{}^P$.

Finally, we study the effect of replacing $\mathcal{O}_{P}$ with its shadow.  To do so, we rearrange \eqref{shadowinverse} and use it to substitute for $\mathcal{O}_{P}$ in \eqref{def_OPEblock}: 
\begin{equation}
    \begin{split}
		\mathcal{O}_{1}(z_1, \bar{z}_1)  \mathcal{O}_{2}(z_2, \bar{z}_2) \Big|_{P}
            &= C'_{12}{}^P\frac{1}{\mathcal{R}_P} \frac{  \Gamma (2-2\bar{h}_P)\Gamma (2 \bar{h}_P) }{\Gamma (2h_P-1)\Gamma(1-2h_P)} \\&  \quad \times \int \frac{d^2 w}{2 \pi} \frac{d^2 z}{2\pi}
           ~ \llangle \mathcal{O}_1(z_1, \bar{z}_1)\mathcal{O}_2(z_2, \bar{z}_2)\widetilde{\mathcal{O}}_P(z, \bar{z}) \rrangle  \llangle \mathcal{O}_P(z, \bar{z}) \mathcal{O}_P(w, \bar{w}) \rrangle \widetilde{\mathcal{O}}_{P}(w, \bar{w}).
    \end{split}
\end{equation} 
As before, we can use \eqref{identity3} to perform the $z$ integral and find  
\begin{equation}
    \begin{split}
	\mathcal{O}_{1}(z_1,& \bar{z}_1) \mathcal{O}_{2}(z_2, \bar{z}_2) \Big|_{P}\\
		&= C'_{12}{}^P\frac{1}{\mathcal{R}_P} (-1)^{h_P-\bar{h}_P +h_{12} - \bar{h}_{12}}  \frac{  \Gamma (2-2\bar{h}_P) }{\Gamma (2h_P-1) }\frac{\Gamma(h_P +h_{12})\Gamma(h_P- h_{12}) }{\Gamma(1-\bar{h}_P - \bar{h}_{12})\Gamma(1-\bar{h}_P+\bar{h}_{12}) }\\&  \quad \times \int \frac{d^2 w}{2 \pi}  
           ~ \llangle \mathcal{O}_1(z_1, \bar{z}_1)\mathcal{O}_2(z_2, \bar{z}_2)  \mathcal{O}_P(w, \bar{w}) \rrangle \widetilde{\mathcal{O}}_{P}(w, \bar{w}) .
    \end{split}
\end{equation}
The right-hand side is manifestly proportional to the OPE block associated to $\mathcal{O}_1\mathcal{O}_2 \to \widetilde{\mathcal{O}}_P$, so we immediately find
\begin{equation}
    \begin{split}
        C'_{12}{}^{\tilde P} = \frac{C'_{12}{}^P}{\mathcal{R}_P}  (-1)^{h_P-\bar{h}_P +h_{12} - \bar{h}_{12}}  \frac{  \Gamma (2-2\bar{h}_P) }{\Gamma (2h_P-1) }\frac{\Gamma(h_P +h_{12})\Gamma(h_P- h_{12}) }{\Gamma(1-\bar{h}_P - \bar{h}_{12})\Gamma(1-\bar{h}_P+\bar{h}_{12}) }.
    \end{split}
\end{equation}
Accounting for the normalization difference between $C$ and $C'$ in \eqref{C-norm}, we 
find that the shadowed OPE coefficients are related by
\begin{equation} \label{OPEsP}
    C_{12}{}^{\tilde P} = \frac{C_{12}{}^P}{\mathcal{R}_P}(-1)^{h_P-\bar{h}_P +h_{12} - \bar{h}_{12}} \frac{\Gamma(2\bar{h}_P)} 
		{\Gamma(1-2h_P )}  \frac {\Gamma(1-h_P-h_{12})\Gamma(1-h_P+h_{12} )} {\Gamma(\bar{h}_P+\bar{h}_{12})\Gamma(\bar{h}_P-\bar{h}_{12} )}.
\end{equation} 
Finally, fixing $\mathcal{R}$ to be given by \eqref{def_R}, this becomes 
\begin{equation}\label{OPEsPfixed}
    C_{12}{}^{\tilde P}
		=  C_{12}{}^P (-1)^{h_P-\bar{h}_P +h_{12} - \bar{h}_{12}} \frac{\Gamma(2h_P-1)\Gamma(2\bar{h}_P)} 
		{\Gamma(2-2\bar{h}_P)\Gamma(1-2h_P )}  \frac {\Gamma(1-h_P-h_{12})\Gamma(1-h_P+h_{12} )} {\Gamma(\bar{h}_P+\bar{h}_{12})\Gamma(\bar{h}_P-\bar{h}_{12} )},
\end{equation}
where again $C_{12}{}^{ \tilde{\tilde P}} =C_{12}{}^P$.

\subsection{Comparison with Existing Literature}
\label{sec:24}

Our results in the previous subsection are universal formulas relating OPE coefficients and their shadows.  A number of studies have also calculated  particular examples of celestial amplitudes in the shadow primary basis.  Here we summarize how our results are related to those in the literature.  

In \cite{Chang:2022jut}, the authors investigate the operator product expansion between two massless scalars in the shadow celestial primary basis that couple via a bulk interaction to a massive scalar. The OPE coefficient is calculated  first as the coefficient of the three-point correlation function, and then in various cases of higher-point functions that can be reduced to calculations involving a three-point structure. Their result (see for example equation (3.12) of \cite{Chang:2022jut}) is consistent with our general formula. Note that their results are just specific examples of the general equivalence between shadowing the OPE block and shadowing the three-point correlation function. 

In \cite{Chang:2022seh}, the authors compute three-point functions in Yang-Mills, massless $\phi^3$ theory, massless scalar QED, and gravity with minimally coupled massless scalars. In each case, all outgoing particles are in the shadow conformal primary basis.  Based on the previous discussion, since these calculations involve only three-point functions, we might expect to be able to reproduce them with our formulas. That is, we might expect our formulas to determine the coefficients of these amplitudes from the coefficients of the corresponding amplitudes in the Mellin conformal primary basis.  However, in this case our formulas are not applicable because  the Mellin amplitudes are constructed from a different branch of the delta function constraint enforcing momentum conservation than the amplitudes in \cite{Chang:2022seh}.  Specifically, \cite{Chang:2022seh} localizes onto a ``soft'' branch of the constraint in which one of the particles carries zero energy, while the Mellin three-point functions localize onto an ``(anti)-holomorphic collinear'' branch in which the positions of all three particles are constrained.  We expect that three-point celestial amplitudes constructed from the same branch will have coefficients related by our formula.

Finally, \cite{Fan:2021isc,Fan:2023lky,Surubaru:2025qhs,Liu:2025dhh} all calculate four-point tree-level amplitudes involving one particle in the shadow basis and \cite{Bhattacharyya:2025nfp} calculates four-point eikonal amplitudes with one particle in the shadow basis.  However, note that since the shadow operation changes the conformal weight, it does not preserve which primary dominates in the coincident limit (in addition to possibly introducing non-local singularities), so our results are not expected to apply directly as they stand.  

\section{Applications} \label{sec:3}

In this section, we study a few applications of the results from the previous section. We consider the shadow of the OPE between a $U(1)$ current and charged operator in Subsection \ref{sec:3.1} and the shadow of the OPE between a stress tensor and a  primary field operator in Subsection \ref{sec:st}. Finally, in Subsection \ref{sec:3.2}, we consider an application in non-abelian gauge theory.

\subsection{U(1) Current OPE} \label{sec:3.1}
			
The bulk large gauge symmetry in $U(1)$ gauge theories is realized by a $U(1)$ Kac-Moody symmetry on the celestial sphere \cite{Strominger:2013lka, Nande:2017dba, He:2014cra}. In this section, we study the effect of the shadow transformation on OPE coefficients of operators charged under this symmetry.
		 
The Ward identity for large $U(1)$ gauge symmetry implies the following OPE between the soft photon current $J$ and a celestial primary operator of charge $Q_i$:
\begin{equation}\label{JO_ope}
	J(z) \mathcal{O}_i (0) \sim \frac{Q_i}{z}~\mathcal{O}_i (0).
\end{equation}
The OPE coefficient for the leading primary is just its charge,
\begin{equation}\label{joo1}
	C_{J \mathcal{O}_i}{}^{\mathcal{O}_i} = Q_i.
\end{equation}
			
We now use the results of the previous section to determine the shadow of the OPE coefficient between the soft photon current $J$ and $\mathcal{O}_i$. First, we use \eqref{OPEs2} to determine the result of shadowing $\mathcal{O}_i$ on the left-hand side of \eqref{JO_ope}. Letting
\begin{equation}
    (h_1, \bar{h}_1) = (1,0), \quad \quad \quad 
    (h_2, \bar{h}_2) = (h_i+\epsilon, \bar{h}_i+\epsilon), \quad \quad \quad
    (h_P, \bar{h}_P) = (h_i, \bar{h}_i),
\end{equation}
and taking the limit $\epsilon \to 0$, we find
\begin{equation}\label{joo2}
	C_{J \tilde{\mathcal{O}}_i}{}^{ \mathcal{O}_i } = C_{J\mathcal{O}_i}{}^{ \mathcal{O}_i} \frac{\mathcal{R}_i}{2\bar{h}_i-1}.
\end{equation}
Note that the  regulator is needed to work directly with the formulas from the previous section, and is consistent with the assumed condition $h_i - \bar{h}_i \in \mathbb{Z}$. Here the normalization $\mathcal{R}$ is left unfixed as it will not appear in the final result.
			
Next, we determine the result of transforming the right-hand  side by using \eqref{OPEsP} with
\begin{equation}
    (h_1, \bar{h}_1) = (1,0), \quad \quad \quad 
    (h_2, \bar{h}_2) = (1-h_i+\epsilon, 1-\bar{h}_i+\epsilon), \quad \quad \quad
    (h_P, \bar{h}_P) = (h_i, \bar{h}_i).
\end{equation}
Then, we find
\begin{equation}\label{joo3}
    \begin{split}
	C_{J \tilde{\mathcal{O}_i}}{}^{ \tilde{\mathcal{O}}_i}
		=  C_{J \tilde{\mathcal{O}_i}}{}^{ \mathcal{O}_i} ~  \frac{1-2\bar{h}_i }{\mathcal{R}_i} .
    \end{split}
\end{equation}
Combining \eqref{joo2}, \eqref{joo3}, and  \eqref{joo1},   we find
\begin{equation} \label{eq:U1sign}
    \begin{split}
        C_{J \tilde{\mathcal{O}}_i}{}^{  \tilde{\mathcal{O}}_i} = - Q_i .
    \end{split}
\end{equation}
If this shadowed OPE coefficient determined the singularity structure away from other operator insertions in a correlation function, it would contribute  
\begin{equation} \label{eq:ShadowCharge}
J(z) \widetilde{\mathcal{O}}_i (0) \sim - \frac{Q_i}{z}~\widetilde{\mathcal{O}}_i  (0),
\end{equation}
and thus immediately suggests an obstruction to working with a mixture of shadowed and unshadowed operators. In particular, the shadow transformation simply smears the operator $O_i$ and therefore cannot change its charge $Q_i$, but based on \eqref{eq:ShadowCharge} the shadowed operator would appear to contribute charge with an opposite sign to the standard charge conservation constraint derived from contour integrals of $J$. However, given that the shadow transform is non-local and takes the original OPE away from its region of convergence (i.e.~away from other operators) in correlation functions, there is reason to expect that additional, possibly non-local, singularity structures will contribute to correlation functions involving $\widetilde{\mathcal{O}}$. The presence of the sign in \eqref{eq:U1sign} is an indication that these structures, and their overall consistency with charge conservation, must be examined more carefully in correlation functions. As discussed in the introduction, this will be the subject of our upcoming work.   
			
\subsection{Stress Tensor OPE} \label{sec:st}
		
In this section, we study the shadow of the OPE involving a stress tensor. The analysis is largely analogous to the one from the previous section. In particular, we begin with the standard OPE for a stress tensor and primary field operator
\begin{equation}
    T(z) \mathcal{O}_i (0) \sim \frac{h_i}{z^2} \mathcal{O}_{i} (0 )+ \frac{1}{z} \partial  \mathcal{O}_{i} (0 ).
\end{equation}
The OPE coefficient for the primary is 
\begin{equation}\label{too1}
	C_{T \mathcal{O}_i}{}^{ \mathcal{O}_i } = h_i,
\end{equation}
and the other singular term is just its first holomorphic descendant.

In analogy with the previous subsection, we use \eqref{OPEs2} with
\begin{equation}
    (h_1, \bar{h}_1) = (2,0), \quad \quad \quad 
    (h_2, \bar{h}_2) = (h_i+\epsilon, \bar{h}_i+\epsilon), \quad \quad \quad
    (h_P, \bar{h}_P) = (h_i, \bar{h}_i).
\end{equation}
Taking the limit $\epsilon \to 0$, we find
\begin{equation}\label{too2}
	C_{T \tilde{\mathcal{O}}_i}{}^{ \mathcal{O}_i} =  C_{T \mathcal{O}_i}{}^{\mathcal{O}_i}~  \mathcal{R}_i  \frac{2 (1-h_i)}{2\bar{h}_i-1} .
\end{equation}
Then, we evaluate \eqref{OPEsP} with
\begin{equation}
    (h_1, \bar{h}_1) = (2,0), \quad \quad \quad (h_2, \bar{h}_2) = (1-h_i+\epsilon, 1-\bar{h}_i+\epsilon), \quad \quad \quad(h_P, \bar{h}_P) = (h_i, \bar{h}_i),
\end{equation}
and we find
\begin{equation}\label{too3}
    \begin{split}
	C_{T \tilde{\mathcal{O}}_i}{}^{  \tilde{\mathcal{O}}_i}= C_{T\tilde{\mathcal{O}}_i}{}^{\mathcal{O}_i}~ \frac{1}{\mathcal{R}_i} \frac{1-2 \bar{h}_i}{2h_i} .
    \end{split}
\end{equation}
Combining \eqref{too2}, \eqref{too3}, and \eqref{too1}, we find
\begin{equation}
    \begin{split}
	C_{T \tilde{\mathcal{O}}_i}{}^{ \tilde{\mathcal{O}}_i}= - (1-h_i) .
    \end{split}
\end{equation}
By analogous logic as in the previous subsection,  if this shadowed OPE coefficient determined the singularity structure away from other operator insertions in a correlation function, it would contribute 
\begin{equation}\label{TOfinal}
	T(z) \widetilde{\mathcal{O}}_{i} (0) \sim - \left[ \frac{1-h_i}{z^2}\widetilde{\mathcal{O}}_{i} (0)+\frac{1}{z} \partial \widetilde{\mathcal{O}}_{i} (0)  \right],
\end{equation}
and would again immediately suggest an obstruction to working with a combination of shadowed and unshadowed operators. Note that the order $z^{-1}$ term is obtained by keeping the first subleading correction in the expansion \eqref{only_P}.

In celestial amplitudes, the stress tensor itself is an example of a celestial shadow state \cite{Kapec:2016jld}.  In particular, it is constructed by shadow-transforming a $\Delta = 0$ negative helicity celestial primary graviton \cite{Fotopoulos:2019tpe, Adamo:2019ipt}. A general OPE with the sign of form \eqref{TOfinal} would therefore be in contradiction in celestial holography with the standard sign  of a $TT$ OPE. More generally, however, these signs again indicate that there are additional singularity structures that contribute to correlation functions beyond the shadow of the OPE coefficient. The subtle nature of the $TT$ OPE within celestial conformal field theory was articulated and first studied in \cite{Fotopoulos:2019vac}. The full singularity structure of correlation functions involving two stress tensors (constructed from shadows of the subleading soft graviton) was indeed found to be in contradiction with the standard $TT$ OPE in \cite{Banerjee:2022wht}, where a prescription using celestial diamonds \cite{Pasterski:2021fjn,Pasterski:2021dqe} was proposed to modify the stress tensor and recover a standard $T_{mod}T_{mod}$ OPE in correlation functions. We will discuss the relation of this prescription to a more general modified shadow prescription (for any spin) in our upcoming work.
			 
\subsection{Chiral Soft Gluon Current Algebra} \label{sec:3.2}

In this section, we discuss the construction of a chiral current algebra of soft gluons. Following the notation in \cite{Pate:2019lpp}, we denote positive/negative helicity celestial primary gluons of conformal weight $\Delta = h+\bar{h}$ by $O^{\pm a}_\Delta$. Here we focus on the operators associated to outgoing gluons. The OPEs of these operators were found to take the form\footnote{Here we do not include delta-function contributions to the operator product expansion that could give rise to a non-vanishing level under a shadow transformation.} \cite{Fan:2019emx} 
\begin{equation}\label{++ope}
    \begin{split}
		O^{+a}_{\Delta_1}(z_1,\bar{z}_1) O^{+b}_{\Delta_2}(z_2,\bar{z}_2)
			\sim \frac{-i f^{ab}{}_c}{z_{12}} B(\Delta_1 -1, \Delta_2-1) O^{+c}_{\Delta_1+\Delta_2-1}(z_2,\bar{z}_2), 
    \end{split}
\end{equation}
\begin{equation} \label{+-ope}
	\begin{split}
		O^{+ a}_{\Delta_1}(z_1,  \bar{z}_1)O^{- b}_{\Delta_2}(z_2,  \bar{z}_2)
		\sim& \frac{-i f^{ab}{}_c}{z_{12}} B(\Delta_1 -1, \Delta_2+1)  O^{- c}_{\Delta_1+\Delta_2-1}(z_2, \bar{z}_2)\\
		&\quad + \frac{-i f^{ab}{}_c}{\bar{z}_{12}} B(\Delta_1+1, \Delta_2-1)  O^{+ c}_{\Delta_1+\Delta_2-1}(z_2,  \bar{z}_2),
	\end{split}
\end{equation}
\begin{equation}\label{--ope}
    \begin{split}
	O^{- a}_{\Delta_1}(z_1,  \bar{z}_1)O^{- b}_{\Delta_2}(z_2,  \bar{z}_2)
	\sim \frac{-i f^{ab}{}_c}{\bar{z}_{12}} B(\Delta_1 -1, \Delta_2-1)  O^{- c}_{\Delta_1+\Delta_2-1}(z_2,  \bar{z}_2).
    \end{split}
\end{equation}
Soft gluon currents, which generate the non-abelian large gauge symmetry, are  given by \cite{Fan:2019emx,Pate:2019mfs,Nandan:2019jas}
\begin{equation}
	J_z^a   \equiv \lim_{\Delta \to 1} (\Delta-1) O^{+a}_\Delta  , \quad \quad \quad
	J_{\bar{z}} ^a   \equiv \lim_{\Delta \to 1}  (\Delta-1) O^{-a}_\Delta .
\end{equation}
			
While in this limit \eqref{++ope} and \eqref{--ope} reduce to the standard expressions for a non-abelian Kac-Moody current algebra (with zero level), the behavior of \eqref{+-ope} in this limit is more subtle. First, the result depends on the order in which  the limits are taken.  This is the celestial  analog of the ambiguity of double soft limits in momentum space, which has recently been understood physically as describing the moduli space of vacua of the theory \cite{Kapec:2022hih}. Second, even if we were to accept that a prescription is necessary, there does not appear to be a natural choice without any drawbacks. For example, if a sequential limit is prescribed -- to be concrete, suppose the limit on positive helicity gluons is taken first -- then the negative helicity gluons simply do not generate a symmetry since they color-rotate all but positive-helicity soft gluons. Given the role played by different helicity soft photons in the magnetic version of large $U(1)$ gauge symmetry \cite{Nande:2017dba,Strominger:2015bla}, a sequential limit would appear to omit some symmetry. On the other hand, if a simultaneous limit is prescribed, then $J_z^a$ ($J_{\bar{z}} ^a$) is not a (anti-)holomorphic current in the presence of other soft insertions.  In these cases,  standard contour deformation arguments cannot be applied. 		
			
Here we show that upon replacing the negative helicity gluons with their shadows and taking a simultaneous limit, we appear to obtain a chiral Kac-Moody current algebra involving only holomorphic currents.\footnote{A similar analysis was performed for abelian gauge theory in \cite{Nande:2017dba}. However in that case, the current algebra describes a free theory. No non-trivial primaries appear in the OPE and hence additional machinery  is not needed.} As emphasized in the previous subsections, we cannot assume that these terms capture the full singularity structure in correlation functions, so the equations presented below will need to be modified and re-interpreted in our upcoming work \cite{Himwich:2025new}. First, a direct application of \eqref{OPEs1fixed}, \eqref{OPEs2fixed}, and \eqref{OPEsPfixed} to \eqref{+-ope} suggests a contribution of the form
\begin{equation}\label{+-ope_shadow}
	\begin{split} 
		O^{+a}_{\Delta_1}(z_1, \bar{z}_1)\widetilde O^{-b}_{2-\Delta_2}(z_2, \bar{z}_2)
   		&\sim- \frac{ -i  f^{ab}{}_c }{z_{12}^{\Delta_1} \bar{z}_{12}^{\Delta_1-1} }  
				\frac{\Gamma (\Delta_1-1) \Gamma ( \Delta_2+1)}{\Gamma (2-\Delta_1+\Delta_2 )}  \widetilde O_{\Delta_1 -\Delta_2+1}^{ -c } (  z_2, \bar{z}_2)\\
			& \quad \quad\quad + \frac{ -i  f^{ab}{}_c }{z_{12}^{\Delta_2}\bar{z}_{12}^{\Delta_2-1} }   \frac{\Gamma (\Delta_1+1) \Gamma (\Delta_2-1)}{\Gamma (2+\Delta_1-\Delta_2)} O_{\Delta_1 -\Delta_2+1}^{ +c } (  z_2, \bar{z}_2), 
	\end{split}
\end{equation} 
and to \eqref{--ope}  of the form 
\begin{equation} \label{--ope_shadow}
    \begin{split}
		\widetilde O^{-a}_{2-\Delta_1}(z_1, \bar{z}_1) \widetilde O^{-b}_{2-\Delta_2}(z_2, \bar{z}_2) 
		&\sim -\frac{ -i  f^{ab}{}_c }{z_{12}  } B(\Delta_1-1,\Delta_2-1) \widetilde O_{3-\Delta_1  -\Delta_2 }^{-c } (  z_2, \bar{z}_2).
	\end{split}
\end{equation}
For this analysis, we use $\mathcal{R}$ given by \eqref{def_R}.  For generic $\Delta_i$, the leading order $z_{12}$ dependence in \eqref{+-ope_shadow} indicates that a more complete treatment in correlation functions is required. Here we simply focus on the contributions above for soft operators for which $\Delta_i$ are integral-valued and leave the interpretation of how these operators contribute in correlation functions for our future work.  
	
To this end, we consider the following soft operators
\begin{equation}
	J^{a}   \equiv \lim_{\epsilon\to 0^+} \epsilon O^{+a}_{1+\epsilon}  , \quad \quad \quad
	\bar  J ^{a}   \equiv -  \lim_{\epsilon \to 0^+} \epsilon  \widetilde O^{-a}_{1+\epsilon} .
\end{equation}
Taking a simultaneous limit \eqref{++ope}, \eqref{+-ope_shadow}, and \eqref{--ope_shadow} gives the  following chiral current algebra:
\begin{equation}\label{current_final}
	\begin{split}
		J^{a}(z_1)J^{b}(z_2)& \sim \frac{-i f^{ab}{}_c}{z_{12}} J^{c}(z_2), \\
		J^{a}(z_1) \bar J^{b}(z_2)& \sim  \frac{-i f^{ab}{}_c}{z_{12}} 
			\frac{1}{2}\left(J^{c}(z_2) -   \bar J^{c}(z_2)  \right), \\
		\bar J^{a}(z_1) \bar J^{b}(z_2)& \sim -\frac{-i f^{ab}{}_c}{z_{12}} \bar J^{c}(z_2).
	\end{split}
\end{equation}
This algebra has a few puzzling features.  First, it does not satisfy the Jacobi identity.  Interestingly, the general form 
\begin{equation}
	\begin{split}
		J^{a}(z_1)J^{b}(z_2)& \sim \frac{-i f^{ab}{}_c}{z_{12}} J^{c}(z_2), \\
		J^{a}(z_1) \bar J^{b}(z_2)& \sim  \frac{-i f^{ab}{}_c}{z_{12}} 
			\left(A J^{c}(z_2) +   B \bar J^{c}(z_2)  \right), \\
		\bar J^{a}(z_1) \bar J^{b}(z_2)& \sim \frac{-i f^{ab}{}_c}{z_{12}} C \bar J^{c}(z_2).
	\end{split}
\end{equation}
only satisfies the Jacobi identity when $A=C, B = 0$ or $A = 0, B=1$ or $A=B=0$. Notice that the first two choices are schematically of the form we might expect from a sequential prescription for soft limits.  The final choice describes a pair of decoupled Kac-Moody currents. 

Next, consider the ``electric'' and ``magnetic'' parts of the currents, following \cite{Strominger:2015bla}: 
\begin{equation}
    J_E  = \frac{1}{2} \left(J  + \bar{J} \right), \quad \quad \quad 
    J_M  = \frac{1}{2i} \left(J  - \bar{J} \right).
\end{equation}
The original currents transform under electromagnetic duality transformations
\begin{equation}
    \begin{split}
        J_E  \to J_M , \quad \quad \quad J_M  \to - J_E, 
    \end{split}
\end{equation}
as
\begin{equation}
    \begin{split}
        J  \to -i J , \quad \quad \quad \bar{J}  \to i \bar{J} . 
    \end{split}
\end{equation}
Then, OPEs of the schematic form 
\begin{equation}
    \begin{split}
        J J  \sim \frac{e+ig}{z}J , \quad \quad \quad
        J \bar{J}  \sim \frac{e+ig}{z}\bar{J}  +\frac{e-ig}{z}J , \quad \quad \quad
        \bar{J} \bar{J}  \sim \frac{e-ig}{z} \bar{J} 
    \end{split}
\end{equation}
are invariant under electromagnetic duality transformations, where the electric and magnetic charges $e$ and $g$ transform as
\begin{equation}
    e \to g, \quad \quad \quad g \to -e. 
\end{equation}
The reality conditions indicate that \eqref{current_final} only captures the ``electric'' part but we also observe that the ``electric'' charges are not of the form to admit an electromagnetic duality-invariant extension. We defer further interpretation of this algebra and these features until the full prescription for OPEs of shadows is understood.
			
\section{Discussion} 
	
In this paper, we derived the shadow of OPE coefficients for celestial operators by shadowing the net contribution from a primary and its global conformal descendants to  (previously-determined) OPEs of celestial primary operators.  A central focus of this paper was to demonstrate the utility of the OPE block, particularly in the study of celestial shadow states. Here we discuss future directions and implications of our work.  

As discussed in the introduction, the shadow of the OPE block does not reflect the full singularity structure of correlation functions involving shadow operators, which will be further discussed in forthcoming work.  A priori the fact that the shadow transform is non-local inhibits a standard notion of OPE. A central question about shadow operators that remains to be addressed is whether shadowed operators can have a meaningful notion of an operator product expansion. Such an expansion is necessary for the celestial dual to admit a stress tensor with a standard OPE -- again, one such proposal has been put forward for this case in \cite{Banerjee:2022wht} -- as well as to  implement various other proposals within celestial holography. Identifying  a general prescription that endows shadow operators with a local OPE will involve a careful investigation of conformal blocks and projections, along the lines of \cite{Simmons-Duffin:2012juh}, for shadow operators.\footnote{For four points, this will include a discussion of partial waves. Shadows of partial waves have appeared in \cite{Chang:2022jut}.} We anticipate that it will involve a modification of the kernel of the OPE block and/or shadow integral.

In this paper, we have remained in $(3,1)$ signature. It has proven fruitful in the study of celestial amplitudes to work in $(2,2)$ signature spacetimes \cite{Atanasov:2021oyu,Melton:2023bjw,Melton:2023hiq}. In that case, the light-ray transform \cite{Kravchuk:2018htv}  is also of interest \cite{Sharma:2021gcz,Guevara:2021abz,Strominger:2021lvk,Himwich:2021dau,Jorge-Diaz:2022dmy}, and the techniques discussed here have been considered in the context of light-ray operators in\cite{Guevara:2021tvr}. We anticipate that analytic continuation will play an important role in formulating a meaningful notion of an operator product expansion for shadowed operators. We hope that this will shed light on the role of bulk analytic continuation to $(2,2)$ signature from the perspective of the celestial dual.
			
The OPE block and its behavior under shadow transformations may also be useful in the study of the massive conformal primary basis.  While massive conformal primary wavefunctions have been constructed \cite{Pasterski:2016qvg,deBoer:2003vf, Narayanan:2020amh, Iacobacci:2020por, Law:2020tsg,Pasterski:2017kqt}, their amplitudes have remained largely unexplored (except for example in \cite{Pasterski:2016qvg,Lam:2017ofc,Himwich:2023njb,Liu:2024lbs,Liu:2025dhh,Ball:2023ukj,Liu:2024vmx}).  For massive particles,  the shadow transform acts within the set of conformal primary wavefunctions (i.e.~no new wavefunctions are found by performing shadow transformations). Hence, any operator product expansion involving massive particles must reflect this property. In addition, the non-locality of the shadow conformal primary wavefunctions and massive conformal primary wavefunctions has a similar flavor, and indeed it has been demonstrated that there are non-local singularities in celestial correlation functions involving currents and massive operators  \cite{Himwich:2023njb}. A prescription for local OPEs of shadow operators is thus likely to shed light on a prescription for local OPEs of massive operators as well. 

As a final note, most of the discourse suggests that some combination of celestial primary operators and their shadows will ultimately be identified as the fundamental degrees of freedom in the celestial dual and that these degrees of freedom will form a local operator algebra on the boundary.  Nevertheless, the non-local behavior of the shadows of these local operators may still carry some physical significance.  In particular, the relation between bulk inversions and boundary shadow transformations \cite{Jorstad:2023ajr,Chen:2024kuq} suggests that the shadow transformation exchanges bulk UV and IR physics. As such, with a proper understanding of how the physics of bulk RG is stored in the boundary, the non-local behavior of the shadows of local boundary degrees of freedom may contain lessons about the UV completion of gravity in the bulk.

\section*{Acknowledgments} 

We are grateful to Jan Albert, Alfredo Guevara, Dan Kapec, Prahar Mitra, Daniele Pranzetti, Ana Raclariu, and Andrew Strominger for useful conversations. This work was supported by NSF grant 2310633. EH is supported by the Princeton Center for Theoretical Science. 
		
\begin{appendix}

\section{Celestial Amplitude Review} \label{appA:celestial-amplitude-review}

In this appendix, we review the systematic procedure for constructing celestial amplitudes (i.e. scattering amplitudes in highest-weight representations) following \cite{Pasterski:2016qvg,Pasterski:2017kqt}.  As mentioned in the introduction, the procedure involves two basic steps.

The first step is to solve the linearized bulk wave equation 
\begin{equation} \label{bulk-wave}
    \begin{split}
        \left(\partial^2 - m^2\right) \varphi(x) = 0, 
    \end{split}
\end{equation}
where $\partial^2$ is shorthand for the linearized differential wave operator, subject to highest weight conditions
\begin{equation} \label{HW-conditions}
    \begin{split}
       L_{-1} \varphi(x) &= \partial_z \varphi(x), \\ 
       L_{0} \varphi(x) &= \left(h+ z\partial_z \right)\varphi(x),\\
       L_{1} \varphi(x) & = \left(2zh+ z^2\partial_z \right)\varphi(x),
    \end{split}
\end{equation}
along with analogous conditions involving $\bar{L}_{0, \pm 1}$ and $h, z$ replaced by $\bar{h}, \bar{z}$.  Here, the global conformal symmetry of two-dimensions ${\rm SL}(2, \mathbb{C})$ is realized as bulk Lorentz transformations ${\rm SO}(3,1)$ through the identification of dual conformal generators $L_m$ and $\bar{L}_m$ with certain combinations of Lorentz generators 
\begin{equation}
    \mathcal{M}_{\mu\nu} = x_\mu \frac{\partial}{\partial x^\nu}-x_\nu \frac{\partial}{\partial x^\mu}.
\end{equation}
Specifically, 
\begin{equation}
    \begin{split}
        L_{\pm 1}& \equiv \frac{i}{2}\left[  \mathcal{M}_{20}+ i \mathcal{M}_{31}\pm \left(\mathcal{M}_{23}- i \mathcal{M}_{10}\right) \right], \quad \quad \quad L_0 \equiv \frac{i}{2}\left(\mathcal{M}_{12}+ i \mathcal{M}_{03}\right),
    \end{split}
\end{equation}
and 
\begin{equation}
    \bar{L}_m = (L_m)^*.
\end{equation}
The resulting solutions $\varphi_{h, \bar{h}}(x;z, \bar{z})$ are thus labeled by left and right conformal weights $(h, \bar{h})$ and a position $(z, \bar{z})$ on the two-dimensional plane. 

The second step is to use these solutions $\varphi_{h, \bar{h}}(x;z, \bar{z})$ as wavefunctions in the LSZ procedure in place of the standard plane waves. As usual, positive and negative frequency modes of $\varphi$ are used to create asymptotic ``out'' versus ``in'' particles, respectively. This step produces scattering amplitudes that transform like correlation functions of primary field operators and thus can be written suggestively as 
\begin{equation}
    \begin{split}
        \langle \mathcal{O}_1 (z_1, \bar{z}_1) \cdots \mathcal{O}_n (z_n, \bar{z}_n) \rangle  = \left[\prod_{i=1}^n\int d^4 x_i ~ \varphi_i(x_i; z_i, \bar{z}_i)\left(\partial^2_i - m_i^2\right)  
        \right]
    \langle T \left \{\phi_1(x_1) \cdots \phi_n(x_n)\right\}\rangle .
    \end{split}
\end{equation}
Here, for notational brevity, we have subsumed all labels including conformal weights as well as other quantum numbers into an index $i$ appearing on the wavefunctions $\varphi_i$, bulk local operators $\phi_i$, and boundary operators $\mathcal{O}_i$.  

As mentioned in the introduction, one can avoid the use of LSZ by instead transforming the momentum-space amplitudes directly to amplitudes in the highest-weight basis. Specifically, by expanding the highest-weight solutions from step one in the plane wave basis, one can determine the transformation from the plane wave basis to the highest-weight basis.  This is possible because the plane waves and the highest-weight solutions each form a complete basis for the wave equation, as discussed below.  Using this basis transformation, the amplitude involving plane-wave creation and annihilation operators can be transformed directly to one involving highest-weight operators. 

\subsection{Highest-weight Solutions}

The massless wave  equation admits two families of highest-weight solutions.  The first is related to the plane wave solutions by a Mellin transform
\begin{equation}
    \begin{split}
        \varphi_{h, \bar{h}}(x; z, \bar{z}) \sim \int_0^\infty \frac{d \omega}{\omega}~ \omega^\Delta e^{i p \cdot x}
    \end{split}
\end{equation}
where $p^\mu$ is a null momentum parameterized an energy scale $\omega$, a sign $\eta = \pm 1$ to distinguish outgoing from incoming momenta, and a point $(z,\bar{z})$ on the celestial sphere: 
\begin{equation} 
	p^\mu =  \eta\omega \hat q^\mu (z, \bar{z}), \quad \quad \quad 
    \hat q^\mu (z, \bar{z}) = \big(1+z\bar{z}, z+\bar{z}, -i(z-\bar{z}), 1-z\bar{z}\big).
\end{equation} 
This family of solutions has the special property that the 2D position $(z, \bar{z})$ coincides with a spatial direction in 4D, namely the direction of the null momentum.  For $\Delta = 1+ i \lambda$,  $\lambda \in \mathbb{R}$, known as the \textit{principal series}, the celestial primary states were found to span the space of plane wave states  \cite{Pasterski:2017kqt}. See \cite{Kulp:2024scx} for additional discussion of principal series versus highest-weight representations.

The second family of solutions to the massless wave equation also spans the space of plane wave solutions for $\Delta = 1+ i \lambda$,  $\lambda \in \mathbb{R}$ \cite{Pasterski:2017kqt} and is related to the first family by a shadow transformation:
\begin{equation} 
    \begin{split}
        \widetilde{\varphi}_{h, \bar{h}}(x; z ,\bar{z})& \sim  \int \frac{d^2 w}{2 \pi } \frac{\varphi_{1-h,1-\bar{h}}(x; w ,w)}{(z-w)^{2h}(\bar{z}-\bar{w})^{2\bar{h}}}. 
    \end{split}
\end{equation} 
For generalizing to the massive case, it is also helpful to note that this second set, when specialized to scalars, can equivalently be written as
\begin{equation} \label{shadow-secondform}
    \begin{split}
        \widetilde{\varphi}_{h, \bar{h}}(x; z ,\bar{z})& \sim \int \frac{d^3 \vec p}{(2 \pi)^3} \frac{1}{2 p^0} \frac{1}{(p \cdot \hat q (z, \bar{z}))^\Delta} e^{ip \cdot x}.
    \end{split}
\end{equation}
In each form, it is clear that the 2D position $(z, \bar{z})$ no longer has a simple interpretation in terms of four-dimensional physics.  

The existence of both families of solutions is guaranteed by symmetry of the linearized wave massless equation under bulk inversions, which map between the two families (see \cite{Jorstad:2023ajr} for additional discussion of this point). 

On the other hand, the massive wave equation is no longer invariant under bulk inversions and as a result admits only a single family of solutions, which span the space of plane wave  solutions for $\Delta = 1+ i \lambda$,  $\lambda \in \mathbb{R}_{\geq 0}$ \cite{Pasterski:2017kqt}. Bulk inversions act non-trivially within this solution space as 2D shadow transformations mapping $\varphi_{h,\bar{h}}\to \widetilde{\varphi}_{h,\bar{h}}\sim \varphi_{1-h,1-\bar{h}}$. The highest-weight scalar solutions take the same form as \eqref{shadow-secondform}: 
 \begin{equation}
     \begin{split}
          \varphi_{h, \bar{h}}(x; z ,\bar{z})
          & \sim \int \frac{d^3 \vec p}{(2 \pi)^3} \frac{1}{2 p^0} \frac{1}{(p \cdot \hat q (z, \bar{z}))^\Delta} e^{ip \cdot x},
     \end{split}
 \end{equation}
 where the only difference is that $p^\mu$ is timelike as opposed to null. Timelike momenta satisfy $p^2 = -m^2$ and are naturally associated with points on a three-dimensional hyperboloid, or equivalently Euclidean ${\rm AdS}_3$.  Hence the Lorentz-invariant measure can be identified with the measure over ${\rm EAdS}_3$:
 \begin{equation}
     \begin{split}
        \int \frac{d^3 \vec p}{(2 \pi)^3} \frac{1}{2 p^0} \sim \int_{{\rm EAdS}_3} [d\hat p],
     \end{split}
 \end{equation}
 where $\hat{p}^\mu$ is the unit vector
 \begin{equation}
     p^\mu = m \hat p^\mu
 \end{equation}
 and the integration kernel can be identified as the Euclidean ${\rm AdS}_3$ bulk-to-boundary propagator:
\begin{equation}
    \begin{split}
        G_{\Delta}(\hat p; z, \bar{z}) = \frac{1}{(\hat{p} \cdot \hat q (z, \bar{z}))^\Delta}.
    \end{split}
\end{equation}
Again, just like for the second branch of solutions to the massless wave equation, the 2D position $(z, \bar{z})$ is not related to the four-dimensional spacetime in a simple way, and thus inhibits a simple bulk interpretation of OPE limits involving massive celestial operators. 

In the massless case, there are also proposals for modified bases using integer weights.  Integer weights directly capture the soft expansion and the associated symmetries \cite{Guevara:2019ypd}. \cite{Cotler:2023qwh} presented a basis of solutions to the massless scalar Klein-Gordon equation that is complete in the sense of decomposing the Wightman function, and involves both primaries and shadows of integer weight. On the other hand, the discrete basis of \cite{Freidel:2022skz}, which is complete in the sense of reconstructing solutions in the Schwartz space, does not involve shadows. Neither does the basis $\Delta\in \mathbb{R}$ considered by the modified contour prescription in \cite{Mitra:2024ugt}.
		
\section{Useful Integral Identities} \label{appB}
			
In the main text, we repeatedly make use of the conformal integral formulas, which can be found in \cite{Dolan:2011dv}. We assume throughout that $h_i - \bar{h}_i \in \mathbb{Z}$.  At two points, we have 
\begin{equation}\label{integral2}
	\int \frac{d^2 z_2}{2 \pi }\frac{1}{z_{12}^{2h }\bar{z}_{12}^{2\bar{h} }} \frac{1}{z_{23}^{2-2h} \bar{z}_{23}^{2-2\bar{h}} }   
		= 2 \pi \delta^{(2)}(z_{13})     \frac{\Gamma (2\bar{h}-1)\Gamma(1-2\bar{h})  }{  \Gamma (2-2h )\Gamma (2h ) }
		=2 \pi \delta^{(2)}(z_{13})     \frac{\Gamma (2h-1)\Gamma(1-2h)  }{  \Gamma (2-2\bar{h} )\Gamma (2\bar{h} ) }.
\end{equation}
At three points,  we have
\begin{equation}\label{identity3}
	\begin{split}
		\int\frac{d^2z}{2 \pi}\frac{1}{(z-z_1)^{h_1}(z-z_2)^{h_2}(z-z_3)^{h_3}}&\frac{1}{(\bar{z}-\bar{z}_1)^{\bar{h}_1}(\bar{z}-\bar{z}_2)^{\bar{h}_2}(\bar{z}-\bar{z}_3)^{\bar{h}_3}}
				\\& \quad \quad \quad \quad \quad
                = K_{123} z_{12}^{h_3-1} z_{23}^{h_1-1} z_{31}^{h_2-1}
				 		\bar{z}_{12}^{\bar{h}_3-1} \bar{z}_{23}^{\bar{h}_1-1} \bar{z}_{31}^{\bar{h}_2-1},
	\end{split}
\end{equation}
where
\begin{equation}
	K_{123} = \frac{\Gamma(1-h_1)\Gamma(1-h_2)\Gamma(1-h_3)}{\Gamma(\bar{h}_1)\Gamma(\bar{h}_2)\Gamma(\bar{h}_3)}
			= \frac{\Gamma(1-\bar{h}_1)\Gamma(1-\bar{h}_2)\Gamma(1-\bar{h}_3)}{\Gamma(h_1)\Gamma(h_2)\Gamma(h_3)}.
\end{equation}
As for two points, we require
\begin{equation}
	h_1+h_2+h_3 = 2, \quad \quad \quad \bar{h}_1+\bar{h}_2+\bar{h}_3 = 2 .
\end{equation}
A more general two-point formula can be found from \eqref{identity3} by sending $z_3, \bar{z}_3 \to \infty$.  The result is
\begin{equation}\label{2point_gen}
	\begin{split}
		\int\frac{d^2z}{2 \pi} \frac{1}{(z-z_1)^{h_1}(z-z_2)^{h_2} } 
			 \frac{1}{(\bar{z}-\bar{z}_1)^{\bar{h}_1}(\bar{z}-\bar{z}_2)^{\bar{h}_2} } 
		& = (-1 )^{h_2-\bar{h}_2} K_{123} z_{12}^{1-h_1-h_2}   \bar{z}_{12}^{1-\bar{h}_1-\bar{h}_2},
		\end{split}
\end{equation}
where $K_{123}$ is evaluated using $h_3 = 2-h_1-h_2$ and $\bar{h}_3 =2-\bar{h}_1-\bar{h}_2$.
		
\end{appendix}

\bibliography{shadowopeI}

\providecommand{\href}[2]{#2}\begingroup\raggedright\begin{thebibliography}{10}

\bibitem{Barnich:2009se}
G.~Barnich and C.~Troessaert, ``{Symmetries of asymptotically flat 4 dimensional spacetimes at null infinity revisited},'' \href{http://dx.doi.org/10.1103/PhysRevLett.105.111103}{{\em Phys. Rev. Lett.} {\bfseries 105} (2010) 111103}, \href{http://arxiv.org/abs/0909.2617}{{\ttfamily arXiv:0909.2617 [gr-qc]}}.

\bibitem{Barnich:2010eb}
G.~Barnich and C.~Troessaert, ``{Aspects of the BMS/CFT correspondence},'' \href{http://dx.doi.org/10.1007/JHEP05(2010)062}{{\em JHEP} {\bfseries 05} (2010) 062}, \href{http://arxiv.org/abs/1001.1541}{{\ttfamily arXiv:1001.1541 [hep-th]}}.

\bibitem{Barnich:2011mi}
G.~Barnich and C.~Troessaert, ``{BMS charge algebra},'' \href{http://dx.doi.org/10.1007/JHEP12(2011)105}{{\em JHEP} {\bfseries 12} (2011) 105}, \href{http://arxiv.org/abs/1106.0213}{{\ttfamily arXiv:1106.0213 [hep-th]}}.

\bibitem{Kapec:2014opa}
D.~Kapec, V.~Lysov, S.~Pasterski, and A.~Strominger, ``{Semiclassical Virasoro symmetry of the quantum gravity $ \mathcal{S}$-matrix},'' \href{http://dx.doi.org/10.1007/JHEP08(2014)058}{{\em JHEP} {\bfseries 08} (2014) 058}, \href{http://arxiv.org/abs/1406.3312}{{\ttfamily arXiv:1406.3312 [hep-th]}}.

\bibitem{Mccarthy:1972ry}
P.~J.~M. Mccarthy, ``{Asymptotically flat space-times and elementary particles},'' \href{http://dx.doi.org/10.1103/PhysRevLett.29.817}{{\em Phys. Rev. Lett.} {\bfseries 29} (1972) 817--819}.

\bibitem{Bekaert:2024jxs}
X.~Bekaert, L.~Donnay, and Y.~Herfray, ``{BMS particles},'' \href{http://arxiv.org/abs/2412.06002}{{\ttfamily arXiv:2412.06002 [hep-th]}}.

\bibitem{Bekaert:2025kjb}
X.~Bekaert and Y.~Herfray, ``{BMS representations for generic supermomentum},'' \href{http://arxiv.org/abs/2505.05368}{{\ttfamily arXiv:2505.05368 [hep-th]}}.

\bibitem{Ahmad:2025new}
S.~Ali~Ahmad, ``{The building blocks of asymptotically flat spacetimes},'' {\em to appear} .

\bibitem{Pasterski:2016qvg}
S.~Pasterski, S.-H. Shao, and A.~Strominger, ``{Flat Space Amplitudes and Conformal Symmetry of the Celestial Sphere},'' \href{http://dx.doi.org/10.1103/PhysRevD.96.065026}{{\em Phys. Rev. D} {\bfseries 96} no.~6, (2017) 065026}, \href{http://arxiv.org/abs/1701.00049}{{\ttfamily arXiv:1701.00049 [hep-th]}}.

\bibitem{Pasterski:2017kqt}
S.~Pasterski and S.-H. Shao, ``{Conformal basis for flat space amplitudes},'' \href{http://dx.doi.org/10.1103/PhysRevD.96.065022}{{\em Phys. Rev. D} {\bfseries 96} no.~6, (2017) 065022}, \href{http://arxiv.org/abs/1705.01027}{{\ttfamily arXiv:1705.01027 [hep-th]}}.

\bibitem{Pasterski:2017ylz}
S.~Pasterski, S.-H. Shao, and A.~Strominger, ``{Gluon Amplitudes as 2d Conformal Correlators},'' \href{http://dx.doi.org/10.1103/PhysRevD.96.085006}{{\em Phys. Rev. D} {\bfseries 96} no.~8, (2017) 085006}, \href{http://arxiv.org/abs/1706.03917}{{\ttfamily arXiv:1706.03917 [hep-th]}}.

\bibitem{Pasterski:2021raf}
S.~Pasterski, M.~Pate, and A.-M. Raclariu, ``{Celestial Holography},'' in {\em {2022 Snowmass Summer Study}}.
\newblock 11, 2021.
\newblock \href{http://arxiv.org/abs/2111.11392}{{\ttfamily arXiv:2111.11392 [hep-th]}}.

\bibitem{Atanasov:2021oyu}
A.~Atanasov, A.~Ball, W.~Melton, A.-M. Raclariu, and A.~Strominger, ``{(2, 2) Scattering and the celestial torus},'' \href{http://dx.doi.org/10.1007/JHEP07(2021)083}{{\em JHEP} {\bfseries 07} (2021) 083}, \href{http://arxiv.org/abs/2101.09591}{{\ttfamily arXiv:2101.09591 [hep-th]}}.

\bibitem{Melton:2023bjw}
W.~Melton, A.~Sharma, and A.~Strominger, ``{Celestial leaf amplitudes},'' \href{http://dx.doi.org/10.1007/JHEP07(2024)132}{{\em JHEP} {\bfseries 07} (2024) 132}, \href{http://arxiv.org/abs/2312.07820}{{\ttfamily arXiv:2312.07820 [hep-th]}}.

\bibitem{Melton:2023hiq}
W.~Melton, A.~Sharma, and A.~Strominger, ``{Conformal correlators on the Lorentzian torus},'' \href{http://dx.doi.org/10.1103/PhysRevD.109.L101701}{{\em Phys. Rev. D} {\bfseries 109} no.~10, (2024) L101701}, \href{http://arxiv.org/abs/2310.15104}{{\ttfamily arXiv:2310.15104 [hep-th]}}.

\bibitem{Kravchuk:2018htv}
P.~Kravchuk and D.~Simmons-Duffin, ``{Light-ray operators in conformal field theory},'' \href{http://dx.doi.org/10.1007/JHEP11(2018)102}{{\em JHEP} {\bfseries 11} (2018) 102}, \href{http://arxiv.org/abs/1805.00098}{{\ttfamily arXiv:1805.00098 [hep-th]}}.

\bibitem{Guevara:2021abz}
A.~Guevara, E.~Himwich, M.~Pate, and A.~Strominger, ``{Holographic symmetry algebras for gauge theory and gravity},'' \href{http://dx.doi.org/10.1007/JHEP11(2021)152}{{\em JHEP} {\bfseries 11} (2021) 152}, \href{http://arxiv.org/abs/2103.03961}{{\ttfamily arXiv:2103.03961 [hep-th]}}.

\bibitem{Strominger:2021lvk}
A.~Strominger, ``{$w_{1+\infty}$ Algebra and the Celestial Sphere: Infinite Towers of Soft Graviton, Photon, and Gluon Symmetries},'' \href{http://dx.doi.org/10.1103/PhysRevLett.127.221601}{{\em Phys. Rev. Lett.} {\bfseries 127} no.~22, (2021) 221601}, \href{http://arxiv.org/abs/2105.14346}{{\ttfamily arXiv:2105.14346 [hep-th]}}.

\bibitem{Sharma:2021gcz}
A.~Sharma, ``{Ambidextrous light transforms for celestial amplitudes},'' \href{http://dx.doi.org/10.1007/JHEP01(2022)031}{{\em JHEP} {\bfseries 01} (2022) 031}, \href{http://arxiv.org/abs/2107.06250}{{\ttfamily arXiv:2107.06250 [hep-th]}}.

\bibitem{Himwich:2021dau}
E.~Himwich, M.~Pate, and K.~Singh, ``{Celestial operator product expansions and w$_{1+\infty}$ symmetry for all spins},'' \href{http://dx.doi.org/10.1007/JHEP01(2022)080}{{\em JHEP} {\bfseries 01} (2022) 080}, \href{http://arxiv.org/abs/2108.07763}{{\ttfamily arXiv:2108.07763 [hep-th]}}.

\bibitem{Guevara:2021tvr}
A.~Guevara, ``{Celestial OPE blocks},'' \href{http://arxiv.org/abs/2108.12706}{{\ttfamily arXiv:2108.12706 [hep-th]}}.

\bibitem{Jorge-Diaz:2022dmy}
C.~Jorge-Diaz, S.~Pasterski, and A.~Sharma, ``{Celestial amplitudes in an ambidextrous basis},'' \href{http://dx.doi.org/10.1007/JHEP02(2023)155}{{\em JHEP} {\bfseries 02} (2023) 155}, \href{http://arxiv.org/abs/2212.00962}{{\ttfamily arXiv:2212.00962 [hep-th]}}.

\bibitem{Ferrara:1972xe}
S.~Ferrara and G.~Parisi, ``{Conformal covariant correlation functions},'' \href{http://dx.doi.org/10.1016/0550-3213(72)90480-4}{{\em Nucl. Phys. B} {\bfseries 42} (1972) 281--290}.

\bibitem{Ferrara:1972ay}
S.~Ferrara, A.~F. Grillo, and G.~Parisi, ``{Nonequivalence between conformal covariant wilson expansion in euclidean and minkowski space},'' \href{http://dx.doi.org/10.1007/BF02815915}{{\em Lett. Nuovo Cim.} {\bfseries 5S2} (1972) 147--151}.

\bibitem{Ferrara:1972uq}
S.~Ferrara, A.~F. Grillo, G.~Parisi, and R.~Gatto, ``{The shadow operator formalism for conformal algebra. Vacuum expectation values and operator products},'' \href{http://dx.doi.org/10.1007/BF02907130}{{\em Lett. Nuovo Cim.} {\bfseries 4S2} (1972) 115--120}.

\bibitem{Ferrara:1972kab}
S.~Ferrara, A.~F. Grillo, G.~Parisi, and R.~Gatto, ``{Covariant expansion of the conformal four-point function},'' \href{http://dx.doi.org/10.1016/0550-3213(73)90467-7}{{\em Nucl. Phys. B} {\bfseries 49} (1972) 77--98}. [Erratum: Nucl.Phys.B 53, 643--643 (1973)].

\bibitem{Simmons-Duffin:2012juh}
D.~Simmons-Duffin, ``{Projectors, Shadows, and Conformal Blocks},'' \href{http://dx.doi.org/10.1007/JHEP04(2014)146}{{\em JHEP} {\bfseries 04} (2014) 146}, \href{http://arxiv.org/abs/1204.3894}{{\ttfamily arXiv:1204.3894 [hep-th]}}.

\bibitem{Gadde:2017sjg}
A.~Gadde, ``{In search of conformal theories},'' \href{http://arxiv.org/abs/1702.07362}{{\ttfamily arXiv:1702.07362 [hep-th]}}.

\bibitem{Caron-Huot:2017vep}
S.~Caron-Huot, ``{Analyticity in Spin in Conformal Theories},'' \href{http://dx.doi.org/10.1007/JHEP09(2017)078}{{\em JHEP} {\bfseries 09} (2017) 078}, \href{http://arxiv.org/abs/1703.00278}{{\ttfamily arXiv:1703.00278 [hep-th]}}.

\bibitem{Simmons-Duffin:2017nub}
D.~Simmons-Duffin, D.~Stanford, and E.~Witten, ``{A spacetime derivation of the Lorentzian OPE inversion formula},'' \href{http://dx.doi.org/10.1007/JHEP07(2018)085}{{\em JHEP} {\bfseries 07} (2018) 085}, \href{http://arxiv.org/abs/1711.03816}{{\ttfamily arXiv:1711.03816 [hep-th]}}.

\bibitem{Jorstad:2023ajr}
E.~J\o{}rstad, S.~Pasterski, and A.~Sharma, ``{Equating extrapolate dictionaries for massless scattering},'' \href{http://dx.doi.org/10.1007/JHEP02(2024)228}{{\em JHEP} {\bfseries 02} (2024) 228}, \href{http://arxiv.org/abs/2310.02186}{{\ttfamily arXiv:2310.02186 [hep-th]}}.

\bibitem{Chen:2024kuq}
H.~Z. Chen, R.~C. Myers, and A.-M. Raclariu, ``{Entanglement, soft modes, and celestial CFT},'' \href{http://dx.doi.org/10.1007/JHEP04(2025)074}{{\em JHEP} {\bfseries 04} (2025) 074}, \href{http://arxiv.org/abs/2403.13913}{{\ttfamily arXiv:2403.13913 [hep-th]}}.

\bibitem{Fan:2019emx}
W.~Fan, A.~Fotopoulos, and T.~R. Taylor, ``{Soft Limits of Yang-Mills Amplitudes and Conformal Correlators},'' \href{http://dx.doi.org/10.1007/JHEP05(2019)121}{{\em JHEP} {\bfseries 05} (2019) 121}, \href{http://arxiv.org/abs/1903.01676}{{\ttfamily arXiv:1903.01676 [hep-th]}}.

\bibitem{Pate:2019lpp}
M.~Pate, A.-M. Raclariu, A.~Strominger, and E.~Y. Yuan, ``{Celestial Operator Products of Gluons and Gravitons},'' \href{http://dx.doi.org/10.1142/9789811210679_0020}{{\em Roman Jackiw} {\bfseries 20} (August, 2020) 217--244}, \href{http://arxiv.org/abs/1910.07424}{{\ttfamily arXiv:1910.07424 [hep-th]}}.

\bibitem{Strominger:2013lka}
A.~Strominger, ``{Asymptotic Symmetries of Yang-Mills Theory},'' \href{http://dx.doi.org/10.1007/JHEP07(2014)151}{{\em JHEP} {\bfseries 07} (2014) 151}, \href{http://arxiv.org/abs/1308.0589}{{\ttfamily arXiv:1308.0589 [hep-th]}}.

\bibitem{He:2015zea}
T.~He, P.~Mitra, and A.~Strominger, ``{2D Kac-Moody Symmetry of 4D Yang-Mills Theory},'' \href{http://dx.doi.org/10.1007/JHEP10(2016)137}{{\em JHEP} {\bfseries 10} (2016) 137}, \href{http://arxiv.org/abs/1503.02663}{{\ttfamily arXiv:1503.02663 [hep-th]}}.

\bibitem{Kapec:2016jld}
D.~Kapec, P.~Mitra, A.-M. Raclariu, and A.~Strominger, ``{2D Stress Tensor for 4D Gravity},'' \href{http://dx.doi.org/10.1103/PhysRevLett.119.121601}{{\em Phys. Rev. Lett.} {\bfseries 119} no.~12, (2017) 121601}, \href{http://arxiv.org/abs/1609.00282}{{\ttfamily arXiv:1609.00282 [hep-th]}}.

\bibitem{Cheung:2016iub}
C.~Cheung, A.~de~la Fuente, and R.~Sundrum, ``{4D scattering amplitudes and asymptotic symmetries from 2D CFT},'' \href{http://dx.doi.org/10.1007/JHEP01(2017)112}{{\em JHEP} {\bfseries 01} (2017) 112}, \href{http://arxiv.org/abs/1609.00732}{{\ttfamily arXiv:1609.00732 [hep-th]}}.

\bibitem{Banerjee:2022wht}
S.~Banerjee and S.~Pasterski, ``{Revisiting the shadow stress tensor in celestial CFT},'' \href{http://dx.doi.org/10.1007/JHEP04(2023)118}{{\em JHEP} {\bfseries 04} (2023) 118}, \href{http://arxiv.org/abs/2212.00257}{{\ttfamily arXiv:2212.00257 [hep-th]}}.

\bibitem{Himwich:2023njb}
E.~Himwich and M.~Pate, ``{w$_{1+\infty}$ in 4D gravitational scattering},'' \href{http://dx.doi.org/10.1007/JHEP07(2024)180}{{\em JHEP} {\bfseries 07} (2024) 180}, \href{http://arxiv.org/abs/2312.08597}{{\ttfamily arXiv:2312.08597 [hep-th]}}.

\bibitem{Fan:2021isc}
W.~Fan, A.~Fotopoulos, S.~Stieberger, T.~R. Taylor, and B.~Zhu, ``{Conformal blocks from celestial gluon amplitudes},'' \href{http://dx.doi.org/10.1007/JHEP05(2021)170}{{\em JHEP} {\bfseries 05} (2021) 170}, \href{http://arxiv.org/abs/2103.04420}{{\ttfamily arXiv:2103.04420 [hep-th]}}.

\bibitem{Crawley:2021ivb}
E.~Crawley, N.~Miller, S.~A. Narayanan, and A.~Strominger, ``{State-operator correspondence in celestial conformal field theory},'' \href{http://dx.doi.org/10.1007/JHEP09(2021)132}{{\em JHEP} {\bfseries 09} (2021) 132}, \href{http://arxiv.org/abs/2105.00331}{{\ttfamily arXiv:2105.00331 [hep-th]}}.

\bibitem{Cotler:2023qwh}
J.~Cotler, N.~Miller, and A.~Strominger, ``{An integer basis for celestial amplitudes},'' \href{http://dx.doi.org/10.1007/JHEP08(2023)192}{{\em JHEP} {\bfseries 08} (2023) 192}, \href{http://arxiv.org/abs/2302.04905}{{\ttfamily arXiv:2302.04905 [hep-th]}}.

\bibitem{Freidel:2022skz}
L.~Freidel, D.~Pranzetti, and A.-M. Raclariu, ``{A discrete basis for celestial holography},'' \href{http://dx.doi.org/10.1007/JHEP02(2024)176}{{\em JHEP} {\bfseries 02} (2024) 176}, \href{http://arxiv.org/abs/2212.12469}{{\ttfamily arXiv:2212.12469 [hep-th]}}.

\bibitem{Mitra:2024ugt}
P.~Mitra, ``{Celestial Conformal Primaries in Effective Field Theories},'' \href{http://arxiv.org/abs/2402.09256}{{\ttfamily arXiv:2402.09256 [hep-th]}}.

\bibitem{Chang:2022seh}
C.-M. Chang and W.-J. Ma, ``{Missing corner in the sky: massless three-point celestial amplitudes},'' \href{http://dx.doi.org/10.1007/JHEP04(2023)051}{{\em JHEP} {\bfseries 04} (2023) 051}, \href{http://arxiv.org/abs/2212.07025}{{\ttfamily arXiv:2212.07025 [hep-th]}}.

\bibitem{Fan:2023lky}
W.~Fan, ``{Celestial conformal blocks of massless scalars and analytic continuation of the Appell function F$_{1}$},'' \href{http://dx.doi.org/10.1007/JHEP01(2024)145}{{\em JHEP} {\bfseries 01} (2024) 145}, \href{http://arxiv.org/abs/2311.11345}{{\ttfamily arXiv:2311.11345 [hep-th]}}.

\bibitem{Chang:2022jut}
C.-M. Chang, W.~Cui, W.-J. Ma, H.~Shu, and H.~Zou, ``{Shadow celestial amplitudes},'' \href{http://dx.doi.org/10.1007/JHEP02(2023)017}{{\em JHEP} {\bfseries 02} (2023) 017}, \href{http://arxiv.org/abs/2210.04725}{{\ttfamily arXiv:2210.04725 [hep-th]}}.

\bibitem{Surubaru:2025qhs}
I.~Surubaru and B.~Zhu, ``{Conformal blocks from celestial graviton amplitudes},'' \href{http://arxiv.org/abs/2501.05805}{{\ttfamily arXiv:2501.05805 [hep-th]}}.

\bibitem{Bhattacharyya:2025nfp}
A.~Bhattacharyya, S.~Ghosh, and S.~Pal, ``{The Sky Remembers everything: Celestial amplitude, Shadow and OPE in quadratic EFT of gravity},'' \href{http://arxiv.org/abs/2505.02899}{{\ttfamily arXiv:2505.02899 [hep-th]}}.

\bibitem{Liu:2025dhh}
R.~Liu and W.-J. Ma, ``{Amplitude from crossing-symmetric celestial OPE},'' \href{http://arxiv.org/abs/2503.21512}{{\ttfamily arXiv:2503.21512 [hep-th]}}.

\bibitem{Nande:2017dba}
A.~Nande, M.~Pate, and A.~Strominger, ``{Soft Factorization in QED from 2D Kac-Moody Symmetry},'' \href{http://dx.doi.org/10.1007/JHEP02(2018)079}{{\em JHEP} {\bfseries 02} (2018) 079}, \href{http://arxiv.org/abs/1705.00608}{{\ttfamily arXiv:1705.00608 [hep-th]}}.

\bibitem{Hijano:2020szl}
E.~Hijano and D.~Neuenfeld, ``{Soft photon theorems from CFT Ward identites in the flat limit of AdS/CFT},'' \href{http://dx.doi.org/10.1007/JHEP11(2020)009}{{\em JHEP} {\bfseries 11} (2020) 009}, \href{http://arxiv.org/abs/2005.03667}{{\ttfamily arXiv:2005.03667 [hep-th]}}.

\bibitem{Duary:2022pyv}
S.~Duary, E.~Hijano, and M.~Patra, ``{Towards an IR finite S-matrix in the flat limit of AdS/CFT},'' \href{http://arxiv.org/abs/2211.13711}{{\ttfamily arXiv:2211.13711 [hep-th]}}.

\bibitem{deGioia:2023cbd}
L.~P. de~Gioia and A.-M. Raclariu, ``{Celestial sector in CFT: Conformally soft symmetries},'' \href{http://dx.doi.org/10.21468/SciPostPhys.17.1.002}{{\em SciPost Phys.} {\bfseries 17} no.~1, (2024) 002}, \href{http://arxiv.org/abs/2303.10037}{{\ttfamily arXiv:2303.10037 [hep-th]}}.

\bibitem{Sleight:2023ojm}
C.~Sleight and M.~Taronna, ``{Celestial Holography Revisited},'' \href{http://dx.doi.org/10.1103/PhysRevLett.133.241601}{{\em Phys. Rev. Lett.} {\bfseries 133} no.~24, (2024) 241601}, \href{http://arxiv.org/abs/2301.01810}{{\ttfamily arXiv:2301.01810 [hep-th]}}.

\bibitem{Jain:2023fxc}
D.~Jain, S.~Kundu, S.~Minwalla, O.~Parrikar, S.~G. Prabhu, and P.~Shrivastava, ``{The S-matrix and boundary correlators in flat space},'' \href{http://arxiv.org/abs/2311.03443}{{\ttfamily arXiv:2311.03443 [hep-th]}}.

\bibitem{Banerjee:2024yir}
S.~Banerjee, ``{Boundary operators in asymptotically flat space-time},'' \href{http://arxiv.org/abs/2406.06690}{{\ttfamily arXiv:2406.06690 [hep-th]}}.

\bibitem{Kapec:2017gsg}
D.~Kapec and P.~Mitra, ``{A $d$-Dimensional Stress Tensor for Mink$_{d+2}$ Gravity},'' \href{http://dx.doi.org/10.1007/JHEP05(2018)186}{{\em JHEP} {\bfseries 05} (2018) 186}, \href{http://arxiv.org/abs/1711.04371}{{\ttfamily arXiv:1711.04371 [hep-th]}}.

\bibitem{Kapec:2021eug}
D.~Kapec and P.~Mitra, ``{Shadows and soft exchange in celestial CFT},'' \href{http://dx.doi.org/10.1103/PhysRevD.105.026009}{{\em Phys. Rev. D} {\bfseries 105} no.~2, (2022) 026009}, \href{http://arxiv.org/abs/2109.00073}{{\ttfamily arXiv:2109.00073 [hep-th]}}.

\bibitem{Kapec:2022hih}
D.~Kapec, ``{Soft particles and infinite-dimensional geometry},'' \href{http://dx.doi.org/10.1088/1361-6382/ad0514}{{\em Class. Quant. Grav.} {\bfseries 41} no.~1, (2024) 015001}, \href{http://arxiv.org/abs/2210.00606}{{\ttfamily arXiv:2210.00606 [hep-th]}}.

\bibitem{Kapec:2022axw}
D.~Kapec, Y.~T.~A. Law, and S.~A. Narayanan, ``{Soft scalars and the geometry of the space of celestial conformal field theories},'' \href{http://dx.doi.org/10.1103/PhysRevD.107.046024}{{\em Phys. Rev. D} {\bfseries 107} no.~4, (2023) 046024}, \href{http://arxiv.org/abs/2205.10935}{{\ttfamily arXiv:2205.10935 [hep-th]}}.

\bibitem{Narayanan:2024qgb}
S.~A. Narayanan, ``{Marginality from Leading Soft Gluons},'' \href{http://arxiv.org/abs/2407.12521}{{\ttfamily arXiv:2407.12521 [hep-th]}}.

\bibitem{Monteiro:2011pc}
R.~Monteiro and D.~O'Connell, ``{The Kinematic Algebra From the Self-Dual Sector},'' \href{http://dx.doi.org/10.1007/JHEP07(2011)007}{{\em JHEP} {\bfseries 07} (2011) 007}, \href{http://arxiv.org/abs/1105.2565}{{\ttfamily arXiv:1105.2565 [hep-th]}}.

\bibitem{Adamo:2021lrv}
T.~Adamo, L.~Mason, and A.~Sharma, ``{Celestial $w_{1+\infty}$ Symmetries from Twistor Space},'' \href{http://dx.doi.org/10.3842/SIGMA.2022.016}{{\em SIGMA} {\bfseries 18} (2022) 016}, \href{http://arxiv.org/abs/2110.06066}{{\ttfamily arXiv:2110.06066 [hep-th]}}.

\bibitem{Ball:2021tmb}
A.~Ball, S.~A. Narayanan, J.~Salzer, and A.~Strominger, ``{Perturbatively exact w$_{1+\infty}$ asymptotic symmetry of quantum self-dual gravity},'' \href{http://dx.doi.org/10.1007/JHEP01(2022)114}{{\em JHEP} {\bfseries 01} (2022) 114}, \href{http://arxiv.org/abs/2111.10392}{{\ttfamily arXiv:2111.10392 [hep-th]}}.

\bibitem{Kmec:2024nmu}
A.~Kmec, L.~Mason, R.~Ruzziconi, and A.~Yelleshpur~Srikant, ``{Celestial Lw$_{1+\infty}$ charges from a twistor action},'' \href{http://dx.doi.org/10.1007/JHEP10(2024)250}{{\em JHEP} {\bfseries 10} (2024) 250}, \href{http://arxiv.org/abs/2407.04028}{{\ttfamily arXiv:2407.04028 [hep-th]}}.

\bibitem{Costello:2022wso}
K.~Costello and N.~M. Paquette, ``{Celestial holography meets twisted holography: 4d amplitudes from chiral correlators},'' \href{http://dx.doi.org/10.1007/JHEP10(2022)193}{{\em JHEP} {\bfseries 10} (2022) 193}, \href{http://arxiv.org/abs/2201.02595}{{\ttfamily arXiv:2201.02595 [hep-th]}}.

\bibitem{Adamo:2023zeh}
T.~Adamo, W.~Bu, and B.~Zhu, ``{Infrared structures of scattering on self-dual radiative backgrounds},'' \href{http://dx.doi.org/10.1007/JHEP06(2024)076}{{\em JHEP} {\bfseries 06} (2024) 076}, \href{http://arxiv.org/abs/2309.01810}{{\ttfamily arXiv:2309.01810 [hep-th]}}.

\bibitem{Melton:2024jyq}
W.~Melton, A.~Sharma, and A.~Strominger, ``{Soft algebras for leaf amplitudes},'' \href{http://dx.doi.org/10.1007/JHEP07(2024)070}{{\em JHEP} {\bfseries 07} (2024) 070}, \href{http://arxiv.org/abs/2402.04150}{{\ttfamily arXiv:2402.04150 [hep-th]}}.

\bibitem{Czech:2016xec}
B.~Czech, L.~Lamprou, S.~McCandlish, B.~Mosk, and J.~Sully, ``{A Stereoscopic Look into the Bulk},'' \href{http://dx.doi.org/10.1007/JHEP07(2016)129}{{\em JHEP} {\bfseries 07} (2016) 129}, \href{http://arxiv.org/abs/1604.03110}{{\ttfamily arXiv:1604.03110 [hep-th]}}.

\bibitem{CarneirodaCunha:2016zmi}
B.~Carneiro~da Cunha and M.~Guica, ``{Exploring the BTZ bulk with boundary conformal blocks},'' \href{http://arxiv.org/abs/1604.07383}{{\ttfamily arXiv:1604.07383 [hep-th]}}.

\bibitem{deBoer:2016pqk}
J.~de~Boer, F.~M. Haehl, M.~P. Heller, and R.~C. Myers, ``{Entanglement, holography and causal diamonds},'' \href{http://dx.doi.org/10.1007/JHEP08(2016)162}{{\em JHEP} {\bfseries 08} (2016) 162}, \href{http://arxiv.org/abs/1606.03307}{{\ttfamily arXiv:1606.03307 [hep-th]}}.

\bibitem{Pate:2020notes}
M.~Pate, ``{Celestial Shadow Operator Product Expansions},'' {\em unpublished} (2020) .

\bibitem{Kulp:2024scx}
J.~Kulp and S.~Pasterski, ``{Multiparticle States for the Flat Hologram},'' \href{http://arxiv.org/abs/2501.00462}{{\ttfamily arXiv:2501.00462 [hep-th]}}.

\bibitem{Himwich:2025new}
E.~Himwich and M.~Pate, ``{Local Celestial Shadow Operators},'' {\em in progress} .

\bibitem{Chang:2018iay}
C.-M. Chang, Y.-H. Lin, S.-H. Shao, Y.~Wang, and X.~Yin, ``{Topological Defect Lines and Renormalization Group Flows in Two Dimensions},'' \href{http://dx.doi.org/10.1007/JHEP01(2019)026}{{\em JHEP} {\bfseries 01} (2019) 026}, \href{http://arxiv.org/abs/1802.04445}{{\ttfamily arXiv:1802.04445 [hep-th]}}.

\bibitem{He:2014cra}
T.~He, P.~Mitra, A.~P. Porfyriadis, and A.~Strominger, ``{New Symmetries of Massless QED},'' \href{http://dx.doi.org/10.1007/JHEP10(2014)112}{{\em JHEP} {\bfseries 10} (2014) 112}, \href{http://arxiv.org/abs/1407.3789}{{\ttfamily arXiv:1407.3789 [hep-th]}}.

\bibitem{Fotopoulos:2019tpe}
A.~Fotopoulos and T.~R. Taylor, ``{Primary Fields in Celestial CFT},'' \href{http://dx.doi.org/10.1007/JHEP10(2019)167}{{\em JHEP} {\bfseries 10} (2019) 167}, \href{http://arxiv.org/abs/1906.10149}{{\ttfamily arXiv:1906.10149 [hep-th]}}.

\bibitem{Adamo:2019ipt}
T.~Adamo, L.~Mason, and A.~Sharma, ``{Celestial amplitudes and conformal soft theorems},'' \href{http://dx.doi.org/10.1088/1361-6382/ab42ce}{{\em Class. Quant. Grav.} {\bfseries 36} no.~20, (2019) 205018}, \href{http://arxiv.org/abs/1905.09224}{{\ttfamily arXiv:1905.09224 [hep-th]}}.

\bibitem{Fotopoulos:2019vac}
A.~Fotopoulos, S.~Stieberger, T.~R. Taylor, and B.~Zhu, ``{Extended BMS Algebra of Celestial CFT},'' \href{http://dx.doi.org/10.1007/JHEP03(2020)130}{{\em JHEP} {\bfseries 03} (2020) 130}, \href{http://arxiv.org/abs/1912.10973}{{\ttfamily arXiv:1912.10973 [hep-th]}}.

\bibitem{Pasterski:2021fjn}
S.~Pasterski, A.~Puhm, and E.~Trevisani, ``{Celestial diamonds: conformal multiplets in celestial CFT},'' \href{http://dx.doi.org/10.1007/JHEP11(2021)072}{{\em JHEP} {\bfseries 11} (2021) 072}, \href{http://arxiv.org/abs/2105.03516}{{\ttfamily arXiv:2105.03516 [hep-th]}}.

\bibitem{Pasterski:2021dqe}
S.~Pasterski, A.~Puhm, and E.~Trevisani, ``{Revisiting the conformally soft sector with celestial diamonds},'' \href{http://dx.doi.org/10.1007/JHEP11(2021)143}{{\em JHEP} {\bfseries 11} (2021) 143}, \href{http://arxiv.org/abs/2105.09792}{{\ttfamily arXiv:2105.09792 [hep-th]}}.

\bibitem{Pate:2019mfs}
M.~Pate, A.-M. Raclariu, and A.~Strominger, ``{Conformally Soft Theorem in Gauge Theory},'' \href{http://dx.doi.org/10.1103/PhysRevD.100.085017}{{\em Phys. Rev. D} {\bfseries 100} no.~8, (2019) 085017}, \href{http://arxiv.org/abs/1904.10831}{{\ttfamily arXiv:1904.10831 [hep-th]}}.

\bibitem{Nandan:2019jas}
D.~Nandan, A.~Schreiber, A.~Volovich, and M.~Zlotnikov, ``{Celestial Amplitudes: Conformal Partial Waves and Soft Limits},'' \href{http://dx.doi.org/10.1007/JHEP10(2019)018}{{\em JHEP} {\bfseries 10} (2019) 018}, \href{http://arxiv.org/abs/1904.10940}{{\ttfamily arXiv:1904.10940 [hep-th]}}.

\bibitem{Strominger:2015bla}
A.~Strominger, ``{Magnetic Corrections to the Soft Photon Theorem},'' \href{http://dx.doi.org/10.1103/PhysRevLett.116.031602}{{\em Phys. Rev. Lett.} {\bfseries 116} no.~3, (2016) 031602}, \href{http://arxiv.org/abs/1509.00543}{{\ttfamily arXiv:1509.00543 [hep-th]}}.

\bibitem{deBoer:2003vf}
J.~de~Boer and S.~N. Solodukhin, ``{A Holographic reduction of Minkowski space-time},'' \href{http://dx.doi.org/10.1016/S0550-3213(03)00494-2}{{\em Nucl. Phys. B} {\bfseries 665} (2003) 545--593}, \href{http://arxiv.org/abs/hep-th/0303006}{{\ttfamily arXiv:hep-th/0303006}}.

\bibitem{Narayanan:2020amh}
S.~A. Narayanan, ``{Massive Celestial Fermions},'' \href{http://dx.doi.org/10.1007/JHEP12(2020)074}{{\em JHEP} {\bfseries 12} (2020) 074}, \href{http://arxiv.org/abs/2009.03883}{{\ttfamily arXiv:2009.03883 [hep-th]}}.

\bibitem{Iacobacci:2020por}
L.~Iacobacci and W.~M\"uck, ``{Conformal Primary Basis for Dirac Spinors},'' \href{http://dx.doi.org/10.1103/PhysRevD.102.106025}{{\em Phys. Rev. D} {\bfseries 102} no.~10, (2020) 106025}, \href{http://arxiv.org/abs/2009.02938}{{\ttfamily arXiv:2009.02938 [hep-th]}}.

\bibitem{Law:2020tsg}
Y.~T.~A. Law and M.~Zlotnikov, ``{Massive Spinning Bosons on the Celestial Sphere},'' \href{http://dx.doi.org/10.1007/JHEP06(2020)079}{{\em JHEP} {\bfseries 06} (2020) 079}, \href{http://arxiv.org/abs/2004.04309}{{\ttfamily arXiv:2004.04309 [hep-th]}}.

\bibitem{Lam:2017ofc}
H.~T. Lam and S.-H. Shao, ``{Conformal Basis, Optical Theorem, and the Bulk Point Singularity},'' \href{http://dx.doi.org/10.1103/PhysRevD.98.025020}{{\em Phys. Rev. D} {\bfseries 98} no.~2, (2018) 025020}, \href{http://arxiv.org/abs/1711.06138}{{\ttfamily arXiv:1711.06138 [hep-th]}}.

\bibitem{Liu:2024lbs}
R.~Liu and W.-J. Ma, ``{Massive celestial amplitudes and celestial amplitudes beyond four points},'' \href{http://dx.doi.org/10.1007/JHEP01(2025)180}{{\em JHEP} {\bfseries 01} (2025) 180}, \href{http://arxiv.org/abs/2404.01920}{{\ttfamily arXiv:2404.01920 [hep-th]}}.

\bibitem{Ball:2023ukj}
A.~Ball, S.~De, A.~Yelleshpur~Srikant, and A.~Volovich, ``{Scalar-Graviton Amplitudes and Celestial Holography},'' \href{http://arxiv.org/abs/2310.00520}{{\ttfamily arXiv:2310.00520 [hep-th]}}.

\bibitem{Liu:2024vmx}
R.~Liu and W.-J. Ma, ``{Celestial optical theorem},'' \href{http://dx.doi.org/10.1103/PhysRevD.111.025017}{{\em Phys. Rev. D} {\bfseries 111} no.~2, (2025) 025017}, \href{http://arxiv.org/abs/2404.18898}{{\ttfamily arXiv:2404.18898 [hep-th]}}.

\bibitem{Guevara:2019ypd}
A.~Guevara, ``{Notes on Conformal Soft Theorems and Recursion Relations in Gravity},'' \href{http://arxiv.org/abs/1906.07810}{{\ttfamily arXiv:1906.07810 [hep-th]}}.

\bibitem{Dolan:2011dv}
F.~A. Dolan and H.~Osborn, ``{Conformal Partial Waves: Further Mathematical Results},'' \href{http://arxiv.org/abs/1108.6194}{{\ttfamily arXiv:1108.6194 [hep-th]}}.

\end{thebibliography}\endgroup
\bibliographystyle{utphys}

\end{document}